\documentclass[twocolumn,english,aps,twocolumn,floatfix,showpacs,tightenlines,superscriptaddress,groupedaddress]{revtex4}
\usepackage[]{fontenc}
\usepackage[latin9]{inputenc}
\usepackage{amsmath}
\usepackage{graphicx}
\usepackage{amssymb}

\makeatletter

\newcommand{\lyxdot}{.}

\@ifundefined{textcolor}{}
{%
 \definecolor{BLACK}{gray}{0}
 \definecolor{WHITE}{gray}{1}
 \definecolor{RED}{rgb}{1,0,0}
 \definecolor{GREEN}{rgb}{0,1,0}
 \definecolor{BLUE}{rgb}{0,0,1}
 \definecolor{CYAN}{cmyk}{1,0,0,0}
 \definecolor{MAGENTA}{cmyk}{0,1,0,0}
 \definecolor{YELLOW}{cmyk}{0,0,1,0}
 }

\usepackage{hyperref}
\usepackage{txfonts}

\makeatother

\usepackage{babel}

\makeatother

\usepackage{babel}

\begin{document}

\title{Unitary equilibrations: probability distribution of the Loschmidt
echo}

\author{Lorenzo Campos Venuti}

\affiliation{Institute for Scientific Interchange, Villa Gualino, Viale Settimio
Severo 65, I-10133 Torino, Italy }

\author{Paolo Zanardi}

\affiliation{Department of Physics and Astronomy, and Center for Quantum Information
Science \& Technology, University of Southern California Los Angeles,
CA 90089-0484 (USA) }

\affiliation{Institute for Scientific Interchange, Villa Gualino, Viale Settimio
Severo 65, I-10133 Torino, Italy }
\begin{abstract}
Closed quantum systems evolve unitarily and therefore cannot converge
in a strong sense to an equilibrium state starting out from a generic
pure state. Nevertheless for large system size one observes \emph{temporal
typicality}. Namely, for the overwhelming majority of the time instants,
the statistics of observables is practically indistinguishable from
an effective equilibrium one. In this paper we consider the Loschmidt
echo (LE) to study this sort of unitary equilibration after a quench.
We draw several conclusions on general grounds and on the basis of
an exactly-solvable example of a quasi-free system. In particular
we focus on the whole probability distribution of observing a given
value of the LE after waiting a long time. Depending on the interplay
between the initial state and the quench Hamiltonian, we find different
regimes reflecting different equilibration dynamics. When the perturbation
is small and the system is away from criticality the probability distribution
is Gaussian. However close to criticality the distribution function
approaches a double peaked, {}``batman-hood'' shaped, universal
form. 
\end{abstract}
\maketitle

\section{introduction}

A sudden change of the parameters governing the evolution of a closed
quantum many-body system gives typically rise to a complex and fascinating
dynamics. This so-called Hamiltonian \emph{quench} is attracting an
increasing amount of attention \cite{calabrese06,cazalilla06,manmana07,silva08}.
The reason for such an interest is, at least, twofold; in the first
place this out of equilibrium phenomenon has been recently observed
in cold atom systems \cite{newtons_cradle,hoffenberth07}. Secondly,
at a more conceptual level, the equilibration dynamics of a quenched
quantum system plays a role in the very foundations of statistical
mechanics \cite{tasaki98,goldsetin06,reimann2,reimann08,winter1,winter2}.
New insights can be gained on the fundamental question about the emergence
of a thermal behavior in closed quantum systems.

In this paper we study a prototypical dynamical quantity for a quantum
quench: the Loschmidt echo (LE). This quantity is defined by the square
modulus of the scalar product, of the time-evolved, (out of equilibrium)
quantum state with the initial (equilibrium) one e.g., Hamiltonian
ground-state.

In spite of the simplicity of its definition the Loschmidt echo $\mathcal{L}$,
or closely related quantities, convey a great deal of information
in a variety of physical problems; for example $\mathcal{L}$ has
been intensively studied in the context of Fermi edge singularities
in x-ray spectra of metals \cite{schotte69}, quantum chaos \cite{prosen98,pastawski01},
decoherence \cite{PZ-LE06,rossini_base07,rossini07}, and more recently
quantum criticality \cite{PZ07} and out-of-equilibrium fluctuations
\cite{jarzynski97,kurchan00}.

Typically (cfr.~figure \ref{fig:LEt}) the Loschmidt echo rapidly
decays from its maximum value $\mathcal{L}=1$ at $t=0$ and, after
an initial transient starts oscillating erraticaly around the same
well defined value. For finite size systems after a sufficiently long
time a pattern of collapses and revivals is observed due to the almost-periodic
nature of the underlying quantum dynamics. On the other hand for infinite
volume systems the Hamiltonian spectrum generically becomes continuous
and an asymptotic value $\mathcal{L}_{\infty}$ (coinciding with the
average one) is eventually reached.

The main goal here is to investigate the statistical properties of
the Loschmidt echo seen as a random variable over the observation
time interval $[0,T].$ One of the key properties is that a small
variance, by standard probability theory arguments, guarantees that
the overwhelming majority of the time $\mathcal{L}(t)$ sticks very
close to its average value \cite{tasaki98,winter2,reimann08}. This
is the sense in which one can speak about \emph{equilibration} dynamics
and corresponding {}``equilibrium properties'' of a finite system
that is evolving unitarily and therefore cannot have attractive fixed
points.

We shall show how the features of the probability distribution of
the Loschmidt echo depend on a rich interplay between the initial
state and quench Hamiltonian on the one hand and the system's size
and observation time on the other. In particular we will focus on
the potential role that the vicinity of quantum critical points may
have on the features of the Loschmidt echo probability distribution
function \cite{PZ07,rossini_base07,silva08}. This latter analysis
will be mostly carried over by exploiting exact results for quasi-free
spin chain i.e., the quantum Ising model \cite{PZ07}.

The paper is organized as follows: in Section \ref{sub:General behavior}
we give the general setting. Later we introduce a relaxation time
$T_{R}$ and discuss the universality content of the \emph{$\mathcal{L}\left(t\right)$
}before this time-scale. In Section \ref{sub:Equilibration-and-long}
we define and study other relevant time scales, the time $T_{1}$
for necessary for observing the correct average, and revival times
where large portion of $\mathcal{L}\left(t\right)$ are back in phase.
In section \ref{sub:Moments-of-the} we give explicit formulas for
the moments of the LE assuming the non-resonant hypothesis. In section
\ref{sec:Ising-Model} we concentrate on a particular example and
prove all the general results advocated so far for an exactly solvable
case. Moreover we discover three universal behaviors for the whole
LE probability distribution function. We draw some parallels with
another natural quenched observable: the magnetization. Finally section
\ref{sec:Conclusions} is devoted to conclusions and outlook.

\section{General Behavior\label{sub:General behavior}}

Let us start by recalling a few elementary yet crucial facts. If $H=\sum_{n}E_{n}\Pi_{n}$
is the system's Hamiltonian ($\Pi_{n}$'s=spectral projections) the
closed-system dynamics is described by the time-evolution superoperator
${\cal U}_{t}=e^{-it{\cal H}},\,{\cal H}(X)=[H,X].$ This superoperator
is thought here of as a map of the space of trace-class operators
$X$ into itself ($\|X\|_{1}:={\rm {tr}}\sqrt{X^{\dagger}X}<\infty$).
Closed quantum systems cannot equilibrate in the strong sense, as
unitary evolutions ${\mathcal{U}}_{t}$ do not have non-trivial i.e.,
non fixed, limit points in the norm topology for $t\rightarrow\infty$
\footnote{Here below we sketch why this is so. Let us suppose that $\rho_{\infty}=\lim_{t\to\infty}{\cal U}_{t}(\rho_{0}).$
Obviously $\rho_{\infty}$ is a fixed point for ${\cal U}_{t}$ i.e.,
${\cal U}_{t}(\rho_{\infty})={\cal U}_{t}(\lim_{u\to\infty}{\cal U}_{u}(\rho_{0}))=\lim_{u\to\infty}{\cal U}_{t+u}(\rho_{0})=\lim_{u\to\infty}{\cal U}_{u}(\rho_{0})=\rho_{\infty}.$
Using unitary invariance of the trace-norm it follows that $\|{\cal U}_{t}(\rho_{0})-\rho_{\infty}\|_{1}=\|{\cal U}_{t}(\rho_{0}-\rho_{\infty})\|_{1}=\|\rho_{0}-\rho_{\infty}\|_{1}$
and therefore $\lim_{t\to\infty}\|{\cal U}_{t}(\rho_{0})-\rho_{\infty}\|_{1}=0$
implies $\rho_{0}=\rho_{\infty}.$%
}. One may then wonder whether a weaker form of convergence can be
achieved for $t\to\infty$.

Let us then consider the expectation value of an observable $A(t):={\rm {tr}\left({\cal U}_{t}(\rho_{0})A\right)}$
and write the spectral resolution of the superoperator ${\cal U}_{t}$
as a formal sum ${\cal U}_{t}=\sum_{{\cal E}}e^{-it{\cal E}}|{\cal E}\rangle\rangle\langle\langle{\cal E}|,$
here ${\cal E}$ ($|{\cal E}\rangle\rangle$)denote the eigenvalues
(eigenvector) of ${\cal H}.$ In finite dimensions the kernel of ${\cal H}$
is spanned by the $\Pi_{n}$'s and gives rise to a time-independent
contribution to $A(t)$ i.e., $A_{\infty}:=\sum_{n}{\rm {tr}}(\Pi_{n}\rho_{0}\Pi_{n}A);$
the point is now to understand whether the remaining components involving
the non-trivial time-dependent factors $\exp(-i{\cal E}t)\,({\cal E}\neq0)$
admits a limit for $t\rightarrow\infty.$ In finite dimensions the
${\cal E}$ are a (finite) discrete set of differences of Hamiltonian
eigenvalues e.g., ${\cal E}=E_{n}-E_{m},$ and correspondingly $A(t)-A_{\infty}$
is a quasi-periodic function: {\em the long time limit of $A(t)$
does not exist.} On the other hand in the infinite dimensional case
the spectrum of ${\cal H}$ can be continuous and, if the function
$\hat{A}({\cal E}):=\langle\rho_{0},{\cal E}\rangle\langle{\cal E},A\rangle$
is sufficiently well behaved, using the Reimann-Lebesgue lemma, $\lim_{t\to\infty}\hat{A}({\cal E})\exp(-i{\cal E}t)=0.$
Therefore in this case \begin{equation}
\lim_{t\to\infty}A(t)=\overline{A(t)}=A_{\infty}\label{lim-inf}\end{equation}
 where $\overline{A(t)}:=\lim_{T\to\infty}\frac{1}{T}\int_{0}^{T}A(t)dt$
denotes the time-average over an infinite time interval. While this
convergence cannot be achieved uniformly for the {\em whole} set
of system's observables (in that it would imply strong convergence)
it can be proven for specific families of $A$'s e.g., local ones
\cite{cramer08}.

There is a third form of convergence that one can consider here: {\em
the convergence in probability}. In the following we shall consider
the above defined $A(t)$ as a random variable over the real line
of $t$ endowed with the uniform measure $dt/T$ with $T\rightarrow\infty.$
Suppose we have a sequence of $\{B_{L}\}_{L}$ (think of $L$ as the
system size), we say that the $B_{L}$'s converge to zero in probability
if \begin{equation}
\lim_{L\to\infty}{\rm {Pr}}\{t\in\mathbf{R}\,/\,|B_{L}(t)|\ge\epsilon\}=0,\,\forall\epsilon>0.\label{prob_conv}\end{equation}
 The meaning of this type of convergence should be clear: for large
$L$ the probability of observing a value of $B_{L}(t)$ different
from zero is vanishingly small. In other words the fractions of $t$'s
for which $B_{L}(t)\neq0$ is going to zero for $L\rightarrow\infty.$
As a matter of fact this is the type of convergence, with $B_{L}(t)=A_{L}(t)-A_{L,\infty},$
that has been considered in \cite{reimann08,winter2}. The stochastic
convergence (\ref{prob_conv}) implies that the probability distributions
of the random variables $A_{L}$ i.e., $P_{L}(\alpha):=\overline{\delta(\alpha-A_{L}(t))}$
are converging to the one of $A_{\infty}:=\lim_{L\to\infty}A_{L,\infty}$
i.e., $\lim_{L\to\infty}P_{L}(\alpha)=\delta(\alpha-A_{\infty}).$
{\em {In this context we say that the initial state $\rho_{0}$
\char`\"{}relaxes\char`\"{} or \char`\"{}equilibrates\char`\"{} to
$\rho_{eq}$ if it happens that $A_{\infty}={\rm {tr}}(A\rho_{eq}).$
}} %
\footnote{Notice that in the infinite-dimensional case discussed above Eq. (\ref{lim-inf})
implies $P(\alpha)=\delta(\alpha-A_{\infty}).$ Indeed from (\ref{lim-inf})
it follows for any continuous $f$ that $\overline{f(A(t))}=f(A_{\infty}).$
Whence $P(\alpha)=\overline{\delta(\alpha-A(t))}=\overline{1/2\pi\int e^{i\lambda\alpha}e^{-i\lambda A(t)}}=1/2\pi\int e^{i\lambda\alpha}\overline{e^{-i\lambda A(t)}}=1/2\pi\int e^{i\lambda\alpha}e^{-i\lambda A_{\infty}}=\delta(\alpha-A_{\infty})$. %
}.

%This  is the  sense in which a unitarily evolving system can said to reach an " equilibrium state" $\rho_{eq}.$
A typical strategy to demonstrate this kind of unitary equilibration
is to prove (\ref{prob_conv}) by showing that i) $\overline{B_{L}(t)}=0$
ii) ${\rm {var}}(B_{L})$ goes to zero sufficiently fast for $L\rightarrow\infty.$
If this is the case one can use a basic probability theory result
${\rm {Pr}}\{t\in\mathbf{R}\,/\,|B_{L}(t)-\overline{B_{L}(t)}|\ge\epsilon\}\le{\rm {var}}(B_{L})/\epsilon^{2}.$
Since, under the assumption ii) the RHS of this latter relation can
be made arbitrarily small and given i), equation (\ref{prob_conv})
holds true. 

Notice that yet another way to formulate the kind of convergence (\ref{prob_conv})
is by means of the concept of \emph{typicality \cite{goldsetin06,reimann2}};
the probability of observing a {}``non-typical'' value i.e.~one
that deviates significantly from the mean one becomes negligible in
the large $L$ limit.

\subsection{The Loschmidt echo \label{sub:The Loschmidt echo}}

\begin{figure}
\noindent \begin{centering}
\includegraphics[width=7cm]{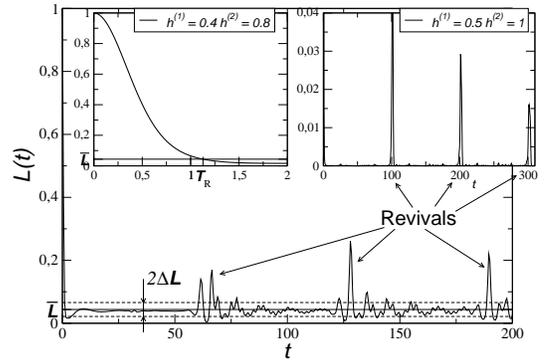} 
\par\end{centering}

\caption{Typical behavior of the Loschmidt echo for the Ising model in transverse
field. All curves refer to a size of $L=100$. In the upper left panel
the relaxation time $T_{R}$ is indicated. Using arguments as in section
\ref{sub:Spectral-analysis} (see also notes \cite{nota,nota-lambda})
one is able to show that, in the quantum Ising model at criticality,
the first revival time is exactly $T_{\mathrm{rev}}=L$ (upper left
panel). \label{fig:LEt}}

\end{figure}

The time dependent quantity we are going to focus on in the rest of
this paper is the Loschmidt echo (LE): \begin{equation}
\mathcal{L}\left(t\right)=\left|\langle\psi|e^{-itH}|\psi\rangle\right|^{2},\label{eq:LE_prima}\end{equation}
 where the state $|\psi\rangle$ is possibly, but not necessarily,
the ground state of the Hamiltonian $H$ at a different coupling.
In the sequel, statistical averages are always taken with respect
to this state. 

In the following we will consider $\mathcal{L}\left(t\right)$ as
a random variable with uniformly distributed $t\ge0$. Ideally we
are interested not only in the first moment but in the whole probability
distribution function. The probability of $\mathcal{L}$ to have value
in $\Omega$ is given by $P\left(\mathcal{L}\in\Omega\right)=\lim_{T\to\infty}T^{-1}\mu\left(\mathcal{L}^{-1}\left(\Omega\right)\cap\left[0,T\right]\right)$.
For those $x$ for which the probability density is well defined,
it is given by\[
P\left(x\right)=\overline{\delta\left(x-\mathcal{L}\left(t\right)\right)}=\lim_{T\to\infty}\frac{1}{T}\sum_{\stackrel{0<t_{n}<T}{\mathcal{L}\left(t_{n}\right)=x}}\frac{1}{\left|\frac{d\mathcal{L}}{dt}\left(t_{n}\right)\right|}.\]

The $k$-th moment of this probability distribution is given by $\mu_{k}:=\int x^{k}P(x)dx=\overline{\mathcal{L}^{k}(t)}.$
Notice that the Loschmidt echo can be written as $\mathcal{L}(t)=\langle\rho_{\psi},e^{-i{\cal H}t}(\rho_{\psi})\rangle,$
where: $\rho_{\psi}:=|\psi\rangle\langle\psi|,\,\mathcal{H}(X):=[H,X]$
and $\langle X,Y\rangle:={\rm {tr}(X^{\dagger}Y)}$ denotes the Hilbert-Schmidt
scalar product. From this it follows $\mathcal{L}^{n}(t)=\langle\rho_{\psi}^{\otimes\, n},e^{-i\mathcal{H}^{(n)}t}(\rho_{\psi}^{\otimes\, n})\rangle,$
where $\mathcal{H}^{(n)}:=\sum_{i=1}^{n}\openone^{\otimes\,(i-1)}\otimes\mathcal{H}\otimes\openone^{\otimes\,(n-i)}.$
Performing the time average %
\footnote{$\mathcal{L}^{n}\left(t\right)=\langle\rho_{\psi}^{\otimes\, n},\mathcal{P}^{\left(n\right)}(\rho_{\psi}^{\otimes\, n})\rangle+\langle\rho_{\psi}^{\otimes\, n},e^{-it\tilde{\mathcal{H}}^{\left(n\right)}}(\rho_{\psi}^{\otimes\, n})\rangle$
where $\tilde{\mathcal{H}}^{\left(n\right)}=\left(\openone-\mathcal{P}^{\left(n\right)}\right)\mathcal{H}^{\left(n\right)}\left(\openone-\mathcal{P}^{\left(n\right)}\right)$.
The time average of the second term is $\langle\rho_{\psi}^{\otimes\, n},F_{T}^{n}(\rho_{\psi}^{\otimes\, n})\rangle/T$
where $F_{T}^{n}=\left(-i\tilde{\mathcal{H}}^{\left(n\right)}\right)^{-1}\left[e^{-iT\tilde{\mathcal{H}}^{\left(n\right)}}-\openone\right]$.
Since $F_{T}^{n}$ is a bounded operator its expectation value divided
by $T$ goes to zero when $T\to\infty$. %
} one finds $\mu_{n}=\langle\rho_{\psi}^{\otimes\, n},{\cal P}^{(n)}(\rho_{\psi}^{\otimes\, n})\rangle,$
where ${\cal P}^{(n)}$ projects onto the kernel of ${\cal H}^{(n)}.$
In particular the time average $\overline{\mathcal{L}}=\mu_{1}$ is
given by \begin{equation}
\overline{\mathcal{L}}=\langle\rho_{\psi},{\cal P}^{(1)}(\rho_{\psi})\rangle=\langle{\cal P}^{(1)}(\rho_{\psi}),{\cal P}^{(1)}(\rho_{\psi})\rangle={\rm {tr}(\rho_{eq}^{2})}\label{eq:LE-average}\end{equation}
 where $\rho_{eq}:={\cal P}^{(1)}(\rho_{\psi}).$ From the general
discussion in Sect. (\ref{sub:General behavior}) we know that ${\cal P}^{(1)}(X)=\sum_{n}\Pi_{n}X\Pi_{n}.$
The effective equilibrium state $\rho_{eq}={\cal P}^{(1)}(\rho_{0})$
is just the $\rho_{0}$ totally dephased in the $H$-eigenbasis.

\subsection{Short time regime and criticality\label{sub:Short-time-regime}}

As already pointed out, typically the LE decays from its maximum value
$1$ at $t=0$ and, after an initial transient, starts oscillating
erraticaly around its mean value. In this section we will analyze
the universality content of this initial transient and its dependence
on the interplay between the initial state $|\psi\rangle$ and the
evolving Hamiltonian $H$.

We start by noticing that the LE Eq.~(\ref{eq:LE_prima}) is the
square modulus of a characteristic function $\chi\left(t\right)=\langle e^{-itH}\rangle$
which is the Fourier transform of the energy probability distribution:
$\hat{\chi}\left(\omega\right)\equiv\langle\delta\left(H-\omega\right)\rangle$.
Both $\chi$ and the LE can be expressed in terms of the cumulants
of $H$:\begin{eqnarray}
\chi\left(t\right) & = & \exp\sum_{n=1}^{\infty}\frac{\left(-it\right)^{n}}{n!}\left\langle H^{n}\right\rangle _{c}\\
\mathcal{L}\left(t\right) & = & \exp2\sum_{n=1}^{\infty}\frac{\left(-t^{2}\right)^{n}}{\left(2n\right)!}\left\langle H^{2n}\right\rangle _{c},\label{eq:losch_cum}\end{eqnarray}
 where $\left\langle \cdot\right\rangle _{c}$ stands for the connected
average with respect to $|\psi\rangle$. The sums above starts from
$n=1$ because the zero order cumulant is zero: $\left\langle H^{0}\right\rangle _{c}=0$.
Since $H$ is a local operator, i.e.~a sum of local {}``variables'',
we can expect in some circumstances, the central limit theorem (CLT)
to apply. More specifically the version of the CLT we are going to
consider here, is the following. \emph{In the thermodynamic limit,
the probability distribution of the rescaled variable $Y\equiv\left(H-\left\langle H\right\rangle \right)/\sqrt{\left\langle H^{2}\right\rangle _{c}}$
tends to a Gaussian (with variance 1 and mean zero) . In other words,
all but the second connected moments of $Y$ tend to zero when the
volume goes to infinity.}

\begin{figure}
\noindent \begin{centering}
\includegraphics[width=70mm]{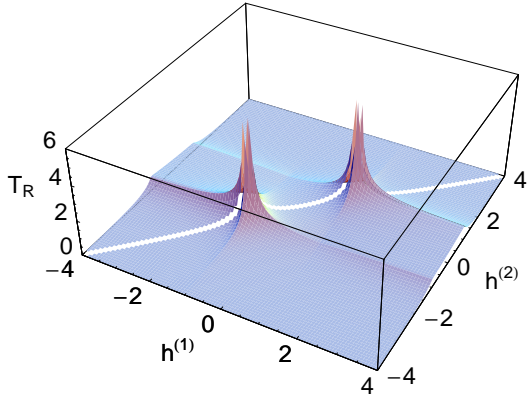} 
\par\end{centering}

\caption{(Color online) Relaxation time for the Loschmidt echo in the Ising
model in transverse field. The critical points are at $h_{c}=\pm1.$Clearly
we observe divergences at critical points when $\delta h$ is small.
On the line $h^{\left(1\right)}=h^{\left(2\right)}$ $\mathcal{L}\left(t\right)=1$
and so there is no relaxation or even dynamics.\label{fig:relax}}

\end{figure}

When the CLT applies, for sufficiently large system sizes, the distribution
of $H$ will be of the form\[
\tilde{\chi}\left(\omega\right)=\frac{1}{\sqrt{2\pi\sigma^{2}}}\exp\left[-\frac{\left(\omega-\left\langle H\right\rangle \right)^{2}}{2\sigma^{2}}\right],\quad\sigma^{2}\equiv\left\langle H^{2}\right\rangle _{c}.\]
 One can systematically compute corrections to this formula and order
them as inverse powers of the system size $L$. Fourier transforming
back we obtain the characteristic function and the LE\begin{equation}
\chi\left(t\right)=e^{it\left\langle H\right\rangle }e^{-\frac{1}{2}t^{2}\sigma^{2}}\Longrightarrow\mathcal{L}\left(t\right)=e^{-t^{2}\sigma^{2}}.\label{eq:LE_CLT}\end{equation}
 It may seem that this expression for the LE could have be obtained
right away by keeping only the first term in the expansion of the
exponential in Eq.~(\ref{eq:losch_cum}):\[
\mathcal{L}\left(t\right)\simeq1-t^{2}\sigma^{2}\simeq e^{-t^{2}\sigma^{2}}.\]
 This is simply a quadratic approximation, that does not rely on the
CLT. Its validity requires $t^{2}\sigma^{2}\ll1$ that in turn implies
$t\ll1/\sqrt{\left\langle H_{2}\right\rangle }$ (as will be explained
later this means roughly $t\ll L^{-d/2}$ where $d$ is the space
dimension). From figure \ref{fig:LEt} we see that typically $\mathcal{L}\left(t\right)$
decays from 1 and after an initial transient, starts oscillating around
an average value $\overline{\mathcal{L}}$ which will be computed
below. Now, when the CLT applies, equation (\ref{eq:LE_CLT}) can
help us to define this transient or \emph{relaxation} time $T_{R}$
given by $e^{-T_{R}^{2}\sigma^{2}}=\overline{\mathcal{L}}$. Roughly
after this time one starts seeing oscillations in $\mathcal{L}\left(t\right)$.
Since in general (see below) one has \emph{$\overline{\mathcal{L}}\simeq e^{-fL^{d}}$},
and when the CLT applies $\sigma^{2}$ scales like the system volume,
the relaxation time scales as $T_{R}=\sqrt{-\ln\overline{\mathcal{L}}/\sigma^{2}}=O\left(L^{0}\right)$.
The situation is different when one considers small variations of
the parameters $\delta h$. In this limit the average LE is related
to the well studied ground state fidelity $F=\left|\langle\psi^{\left(1\right)}|\psi^{\left(2\right)}\rangle\right|$.
More precisely one has $\overline{\mathcal{L}}\simeq F^{4}$ \cite{rossini07}.
Close to critical points the behavior of the fidelity is dictated
by the scaling dimension $\Delta$ of the most relevant operator in
$H$ with respect to the critical state $|\psi\rangle$ \cite{LCV07}.
The precise scaling is the following: $F\sim1-\mathrm{const.}\times\delta h^{2}L^{2\left(d+\zeta-\Delta\right)}$,
where $\zeta$ is the dynamical critical exponent. Similarly one can
show (see below) that $\sigma^{2}\sim L^{2\left(d-\Delta\right)}$.
All in all this amounts to saying that the relaxation time for small
variation $\delta h$ around a critical point (roughly $\delta h\ll L^{-\left(d+\zeta-\Delta\right)}$)
increases from $O\left(1\right)$ to $T_{R}\sim L^{\zeta}$. In the
thermodynamic limit instead, i.e.~taking first the limit $L\to\infty$,
using standard scaling arguments, one can show that for $\delta h$
small and $h$ close to the critical point $h_{c}$, the relaxation
times diverges as $T_{R}\sim\left|h-h_{c}\right|^{-\zeta\nu}$, $\nu$
being the correlation length exponent. In figure \ref{fig:relax}
we plot the relaxation time for a concrete example that will be studied
thoroughly in section \ref{sec:Ising-Model}, the Ising model in transverse
field. There one has $\Delta=\zeta=\nu=1$, and so the singularities
observed in figure \ref{fig:relax} are of simple algebraic type:
$T_{R}\sim\left|h-h_{c}\right|^{-1}$ around the critical points $h_{c}=\pm1$.

Let us now turn to discuss when we expect the CLT to work. Consider
first the case when $|\psi\rangle$ is the ground state of a gapped
Hamiltonian and the connected energy correlators go to zero exponentially
fast: $\langle H\left(x\right)H\left(y\right)\rangle\stackrel{\left|x-y\right|\to\infty}{\longrightarrow}\langle H\left(x\right)\rangle\langle H\left(y\right)\rangle$
(exponential clustering). In this case the connected averages of $H$
scale as the volume: $\left\langle H^{n}\right\rangle _{c}\sim L^{d}$.
As a consequence the cumulants of the rescaled variable satisfy $\left\langle Y^{n}\right\rangle _{c}\sim L^{-\left(nd-2d\right)/2}$
for $n\ge2$, which immediately implies the CLT in the sense given
above.

Therefore we can have violation of the CLT only in the gapless case
when the state $|\psi\rangle$ is critical or when clustering fails.
Let us then consider a critical state $|\psi\rangle$. Connected averages
have a regular extensive part and a singular part which scales according
to the most relevant component of $H$ with scaling dimension $\Delta$.
When $|\psi\rangle$ is the ground state of $H$ at a different coupling,
$\Delta$ is the scaling dimension of the perturbation $\delta H$
to the critical Hamiltonian. At leading order, we can write $\left\langle H^{n}\right\rangle _{c}\sim A_{n}L^{d}+B_{n}L^{n\left(d-\Delta\right)}$,
and so the rescaled variable satisfies \[
\left\langle Y^{n}\right\rangle _{c}\sim\frac{A_{n}L^{d}+B_{n}L^{n\left(d-\Delta\right)}}{\left(A_{2}L^{d}+B_{2}L^{2\left(d-\Delta\right)}\right)^{n/2}}.\]
 If $\Delta<d/2$ the cumulants of the rescaled variable $Y$ don't
go to zero but to universal constants \cite{privman,aji01} given
by \[
\left\langle Y^{n}\right\rangle _{c}\to\frac{B_{n}}{B_{2}^{n/2}}.\]
 In this case the probability distribution of the energy is a, non-Gaussian,
universal distribution. This kind of universal behavior has been observed
for instance in \cite{fendley08} on an example where the scaling
dimension is $\Delta=1/8$. In the opposite situation where $\Delta>d/2$,
all the cumulants of $Y$ go to zero except for the first two, and
the distribution function approaches a Gaussian in the large size
limit. In the intermediate case $\Delta=d/2$ the cumulants of $Y$
do not go zero but to a constant which is however not universal due
to extensive contributions coming from the denominator. We recall
that in $d$ dimensional, zero temperature quantum mechanics, operators
are classified into relevant, irrelevant, and marginal if their scaling
dimension is respectively smaller, larger, or equal to $d+\zeta$
where $\zeta$ is the dynamical exponent. Hence we see that, even
in the critical case, we observe deviation from the Gaussian behavior
only if the perturbation $\delta H$ is sufficiently relevant, specifically
$\Delta<d/2$.

To finish let us remind the reader that the CLT also breaks down when
clustering fails.

\begin{figure}
\noindent \begin{centering}
\includegraphics[width=70mm]{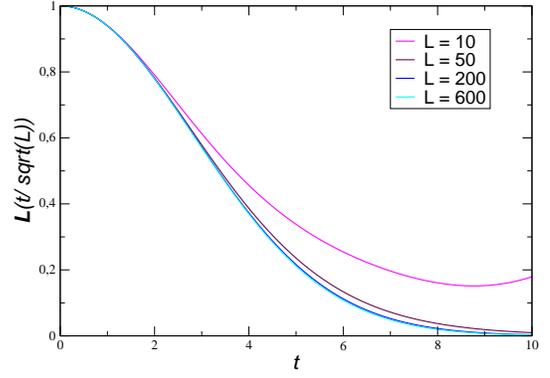} 
\par\end{centering}

\caption{(Color online) Rescaled Loschmidt echo $\mathcal{L}_{L}\left(t/\sqrt{L}\right)$
for the Ising model in transverse field. System sizes are, from top
to bottom, $L=10,50,200,600$. The state which defines the average
is critical: $h^{\left(1\right)}=1$ while $H$ is at $h^{\left(2\right)}=2.5$.
The same data collapse feature is observed when choosing different
$h^{\left(i\right)}$s, although the variance of this Gaussian is
sensitive to that. \label{fig:losch_CLT}}

\end{figure}

To summarize, when the CLT applies, the LE tends to a Gaussian and
plotting the function $\mathcal{L}_{L}\left(t/\sqrt{\left\langle H^{2}\right\rangle _{2}}\right)$
for different sizes $L$ one should observe data collapse (see figure
\ref{fig:losch_CLT}).

\subsection{Equilibration and long time behavior\label{sub:Equilibration-and-long}}

After having discussed the short time behavior of the LE related to
the initial transient, let us now turn to its long time behavior.

We first re-write Eq.~(\ref{eq:LE_prima}) in the eigenbasis of $H=\sum_{n}E_{n}|n\rangle\langle n|$\begin{equation}
\mathcal{L}\left(t\right)=\sum_{n,m}p_{n}p_{m}e^{-it\left(E_{n}-E_{m}\right)},\label{eq:LE-base}\end{equation}
 where $p_{n}=\left|\langle\psi|n\rangle\right|^{2}$.

If the spectrum of $H$ is non degenerate the superoperator ${\cal P}^{(1)}$
acts as a dephasing in the Hamiltonian eigenbasis i.e, ${\cal P}^{(1)}(X)=\sum_{n}\langle n|X|n\rangle|n\rangle\langle n|.$
In other words the time average of the exponentials in Eq.~(\ref{eq:LE-base})
gives simply $\delta_{n,m}$ and equation (\ref{eq:LE-average}) reduces
to $\overline{\mathcal{L}}=\sum_{n}p_{n}^{2}$. As is well known this
quantity is the purity of an equilibrium, dephased, state: $\rho_{\mathrm{eq}}=\sum_{n}p_{n}|n\rangle\langle n|$.

\paragraph*{Time scales}

In the preceding section we already defined a relevant time scale,
the relaxation time $T_{R}$ which is $O\left(1\right)$ off-criticality
while $T_{R}=O\left(L^{\zeta}\right)$ in the critical case and for
sufficiently small variations $\delta h\ll L^{-\left(d+\zeta-\Delta\right)}$.

In some situations it is useful to consider a finite observation time
$T$. We will write $\overline{\mathcal{L}}$ to indicate the corresponding
average. It is natural to ask about the interplay between the observation
time $T$ and the linear size of the system $L$. In other words in
general $\lim_{L\to\infty}\lim_{T\to\infty}\overline{\mathcal{L}_{L}}\neq\lim_{T\to\infty}\lim_{L\to\infty}\overline{\mathcal{L}_{L}}$.
Since taking larger system sizes has the effect of sending the revival
times to infinity and the LE attunes its maximum value $1$ at $t=0$,
typically the function $\lim_{L\to\infty}\mathcal{L}_{L}\left(t\right)$
has only one large peak at $t=0$ whereas $\mathcal{L}_{L}\left(t\right)$
has peaks at all the revival times. Correspondingly we expect $\lim_{L\to\infty}\lim_{T\to\infty}\overline{\mathcal{L}_{L}}>\lim_{T\to\infty}\lim_{L\to\infty}\overline{\mathcal{L}_{L}}$.
This expectation has been confirmed for the case of the one dimensional
quantum Ising model, see section \ref{sec:Ising-Model}.

Another question which is relevant in the measurement process is how
large must the observation time be to effectively measure $\overline{\mathcal{L}}$?
That is, what is the condition to have $\overline{\mathcal{L}^{T_{1}}}=\overline{\mathcal{L}}$
or more in general what is the smallest time $T_{n}$ such that one
observes $\overline{\left[\left(\mathcal{L}\right)^{n}\right]^{T_{n}}}=\overline{\left(\mathcal{L}\right)^{n}}$?
Let us focus on $T_{1}$. Looking at equation (\ref{eq:LE-base})
one realizes that it suffices to have $T\gg\Delta_{\mathrm{min}}^{-1}$,
where $\Delta_{\mathrm{min}}$ is the smallest gap in the whole spectrum
i.e.~$\Delta_{\mathrm{min}}=\min_{n,m}\left(E_{n}-E_{m}\right)$.
We can address this question for the class of quasi-free Fermi systems
in $d$ spatial dimensions. In this case the energy has the form $E_{n}=\sum_{\boldsymbol{k}}n_{\boldsymbol{k}}\Lambda_{\boldsymbol{k}}$
where $\boldsymbol{k}$ is a $d$-dimensional quasi-momentum-like
label and we can assume the one particle energy $\Lambda_{\boldsymbol{k}}$
to be positive. Then the gap is given by $\Delta_{\mathrm{min}}=\min_{\beta_{\boldsymbol{k}}}\left|\sum_{\boldsymbol{k}}\beta_{\boldsymbol{k}}\Lambda_{\boldsymbol{k}}\right|$
with $\beta_{\boldsymbol{k}}=0,\pm1$. By choosing $\beta_{\boldsymbol{k}}=\left(-1\right)^{\boldsymbol{k}}$
, one obtains a $\Delta_{\mathrm{min}}$ which is exponentially small
in $L$ in those (frequent) cases where $\Lambda_{\boldsymbol{k}}$
is an analytic function of $\boldsymbol{k}$ \cite{nota-gap}. However
in quasi free systems the weights $p_{n}$ decrease exponentially
with the number of excitations in $n$. In practice the highest weight
is given for energy differences between the one and zero particle
spectra: $\Delta^{\left(1,0\right)}=\min_{\boldsymbol{k}}\Lambda_{\boldsymbol{k}}$
which is a constant of order 1 in the gapful case, while typically
scales as $L^{-1}$ for the critical case. The next largest amount
of spectral weight is attained at a gap which is a difference between
one particle energies $\Delta^{\left(1,1\right)}=\min_{\boldsymbol{k},\boldsymbol{q}}\left|\Lambda_{\boldsymbol{k}}-\Lambda_{\boldsymbol{q}}\right|$.
It will be favorable to have $\boldsymbol{k}$ and $\boldsymbol{q}$
nearby in the region where $\Lambda_{\boldsymbol{k}}$ is flat or
almost flat. So we get $\Delta^{\left(1,1\right)}=\min_{\boldsymbol{k}}\left|\Lambda_{\boldsymbol{k}}-\Lambda_{\boldsymbol{k}+\delta\boldsymbol{k}}\right|\simeq\min_{\boldsymbol{k}}\left|\nabla_{\boldsymbol{k}}\Lambda_{\boldsymbol{k}}\cdot\delta\boldsymbol{k}\right|$.
This gap is at least of order of $L^{-2}$ (or at least $O\left(L^{-3}\right)$
if there exists a $\boldsymbol{k}$-vector such that $\nabla_{\boldsymbol{k}}\Lambda_{\boldsymbol{k}}=0$).
From this discussion we estimate that, at least in quasi-free systems,
to have $\overline{\mathcal{L}^{T_{1}}}\simeq\overline{\mathcal{L}}$
one must take $T_{1}=O\left(1\right)$ in the gapful case, while one
has $T_{1}=O\left(L\right)$ at criticality. If one needs $\overline{\mathcal{L}^{T_{1}}}\simeq\overline{\mathcal{L}}$
with a larger degree of precision than one must choose considerably
larger time: $T_{1}=O\left(L^{2}\right)$ (or $T_{1}=O\left(L^{3}\right)$
if $\nabla_{\boldsymbol{k}}\Lambda_{\boldsymbol{k}}=0$ has a solution
within the allowed set of $\boldsymbol{k}$-vectors). Related time
scales are \emph{revival times}. We define a revival time to be that
particular time for which a large portion of spectral weight $p_{n}p_{m}$
has revived. More precisely $T_{\mathrm{rev}}\omega_{\mathrm{peak}}=2\pi$,
where $\omega_{\mathrm{peak}}$ is a particular frequency $E_{n}-E_{m}$
such that the weight $p_{n}p_{m}$ is large. From the discussion above
we expect $T_{\mathrm{rev}}=O\left(1\right)$ when $H$ has a gap
above the ground state, while $T_{\mathrm{rev}}=O\left(L\right)$
when $H$ is critical. These expectations have been confirmed (see
figure \ref{fig:LEt}) on the hand of a solvable model that will be
discussed in the next sections.

\subsection{Moments of the Loschmidt echo\label{sub:Moments-of-the}}

Having computed the time averaged LE we can now turn to higher moments.
In doing this one has to distinguish cases where $n=m$ from those
where $E_{n}=E_{m}$ in Eq.~(\ref{eq:LE-base}). So we write\begin{eqnarray*}
\mathcal{L}\left(t\right) & = & \overline{\mathcal{L}}+X\left(t\right),\\
X\left(t\right) & = & \sum_{n\neq m}p_{n}p_{m}e^{-it\left(E_{n}-E_{m}\right)},\end{eqnarray*}
and the $n$-th moment is given by\[
\overline{\left[\mathcal{L}\left(t\right)\right]^{n}}=\sum_{k=0}^{n}\left(\begin{array}{c}
n\\
k\end{array}\right)\overline{\mathcal{L}}^{n-k}\overline{\left[X\left(t\right)\right]^{k}}.\]
 The computation of the average $\overline{\left[X\left(t\right)\right]^{k}}$
can be done assuming a \emph{strong non-resonance} condition. With
this we mean the following. We say $H$ satisfies a $k$-non-resonance
condition if the only way to fulfill $\sum_{l=1}^{k}E_{i_{l}}-E_{j_{l}}=0$
is to match the $E_{i}$'s to the $E_{j}$'s. Strong non resonance
is $k$-non-resonance for any $k$. Note that this condition cannot
be fulfilled when $k$ becomes of the order of the Hilbert's space
dimension. Now to compute $\overline{\left[X\left(t\right)\right]^{k}}$
draw $2k$ points in two rows of length $k$. Imagine $E_{i}$ ($E_{j}$)
are the points at the left (right). Now draw all possible contraction
between $i$'s and $j$'s (no contraction among $i$'s or $j$'s since
they have the same sing), which are $k!$, but \emph{keep only those
sets of contractions where there is no horizontal line}. This requirement
corresponds to the constraint $i_{l}\neq j_{l},\,\, l=1,\ldots,k$.
For example for $\left[X\left(t\right)\right]^{2}$ we have only one
contribution

\noindent \begin{center}
\includegraphics[scale=0.3]{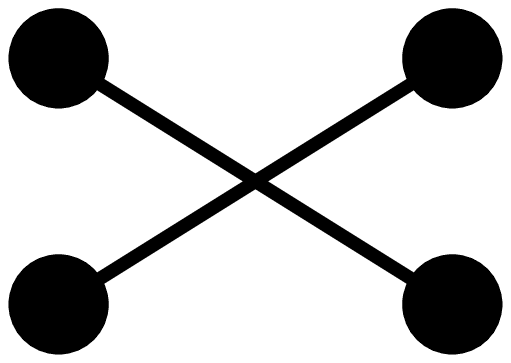} 
\par\end{center}

so\[
\overline{\left[X\left(t\right)\right]^{2}}=\sum_{i_{1}\neq i_{2}}p_{i_{1}}^{2}p_{i_{2}}^{2}\]
 For $\overline{\left[X\left(t\right)\right]^{3}}$ we have two diagrams:

\noindent \begin{center}
\includegraphics[scale=0.3]{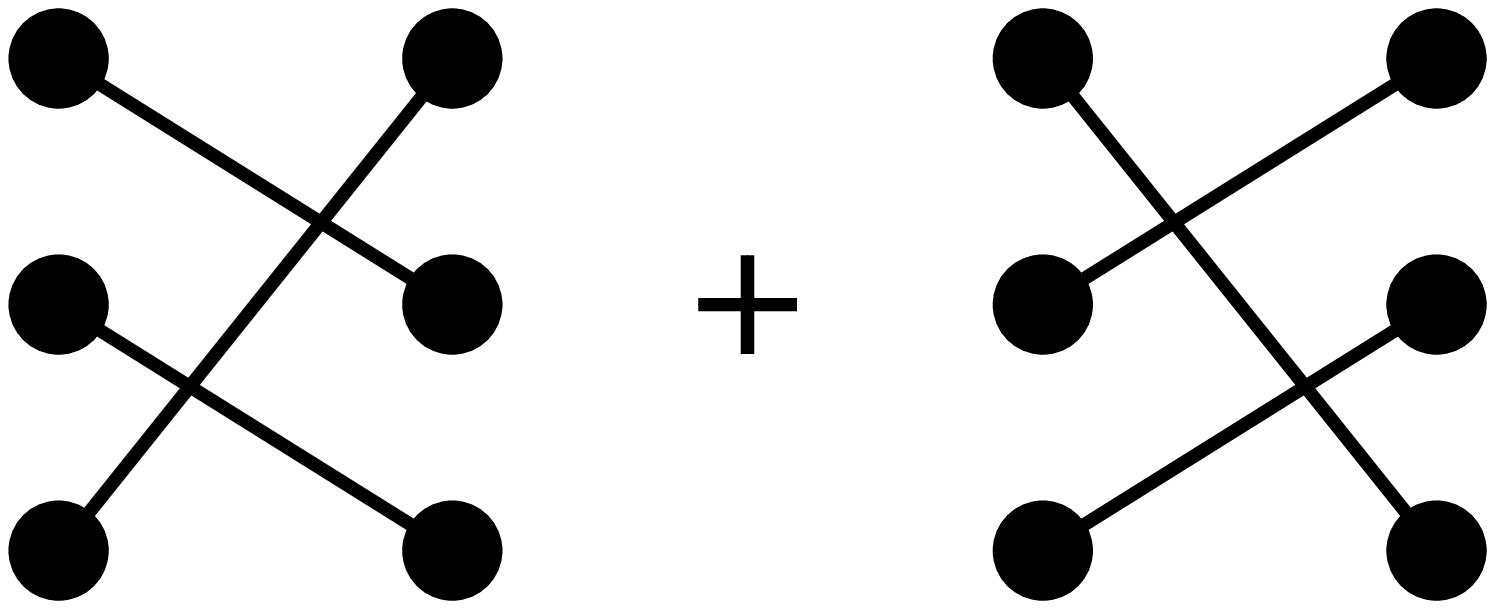} 
\par\end{center}

Both these diagrams give the same contribution (simply swap $i$'s
with $j$'s) and the result is\[
\overline{\left[X\left(t\right)\right]^{3}}=2\sum_{\substack{i_{1}\neq i_{2}\\
i_{2}\neq i_{3},i_{3}\neq i_{1}}
}p_{i_{1}}^{2}p_{i_{2}}^{2}p_{i_{3}}^{2}.\]
 The number of terms in $\overline{\left[X\left(t\right)\right]^{k}}$,
$N\left(k\right)$ is the number of all permutations without fixed
points and is given by\[
N\left(k\right)=\sum_{j=2}^{k}\left(-1\right)^{k-j}\left(\begin{array}{c}
k\\
j\end{array}\right)\left(j!-1\right).\]
 However, among these $N\left(k\right)$ terms, many of them give
different contributions. Look for instance at $\overline{\left[X\left(t\right)\right]^{4}}$:\[
\overline{\left[X\left(t\right)\right]^{4}}=3\left(\sum_{i_{1}\neq i_{2}}p_{i_{1}}^{2}p_{i_{2}}^{2}\right)^{2}+6\sum_{\substack{i_{1}\neq i_{2},i_{2}\neq i_{3}\\
i_{3}\neq i_{4},i_{4}\neq i_{1}}
}p_{i_{1}}^{2}p_{i_{2}}^{2}p_{i_{3}}^{2}p_{i_{4}}^{2}\]
 Correctly one has $3+6=N\left(4\right)=9$.

We collect here the first three moments\begin{eqnarray*}
\mu_{1} & = & \overline{\mathcal{L}}\\
\mu_{2} & = & \overline{\mathcal{L}}^{2}+\sum_{i_{1}\neq i_{2}}p_{i_{1}}^{2}p_{i_{2}}^{2}\\
\mu_{3} & = & \overline{\mathcal{L}}^{3}+3\overline{\mathcal{L}}\sum_{i_{1}\neq i_{2}}p_{i_{1}}^{2}p_{i_{2}}^{2}+2\sum_{\substack{i_{1}\neq i_{2}\\
i_{2}\neq i_{3},i_{3}\neq i_{1}}
}p_{i_{1}}^{2}p_{i_{2}}^{2}p_{i_{3}}^{2}\end{eqnarray*}
 while the cumulants are\begin{eqnarray*}
\kappa_{1} & = & \overline{\mathcal{L}}\\
\kappa_{2} & = & \sum_{i_{1}\neq i_{2}}p_{i_{1}}^{2}p_{i_{2}}^{2}\\
\kappa_{3} & = & 2\sum_{\substack{i_{1}\neq i_{2}\\
i_{2}\neq i_{3},i_{3}\neq i_{1}}
}p_{i_{1}}^{2}p_{i_{2}}^{2}p_{i_{3}}^{2}.\end{eqnarray*}

We can notice that each term in $\overline{\left[X\left(t\right)\right]^{k}}$
has the same form of $\overline{\mathcal{L}}^{k}$ except for a number
of non-resonance constraints of the form $i_{l}\neq i_{m}$. Correspondingly
$\overline{\left[X\left(t\right)\right]^{k}}<N\left(k\right)\overline{\mathcal{L}}^{k}$
. Using now $\sum_{k=0}^{n}\left(\begin{array}{c}
n\\
k\end{array}\right)N\left(k\right)=n!$ we obtain the simple bound $\overline{\mathcal{L}^{n}}<n!\overline{\mathcal{L}}^{n}$
for $n\ge2$. This means that\[
e^{\lambda\overline{\mathcal{L}}}\le\tilde{\chi}\left(\lambda\right)=\overline{e^{\lambda\mathcal{L}}}<\frac{1}{1-\overline{\mathcal{L}}\lambda},\]
 and so we obtained a bound on the characteristic function $\tilde{\chi}$.
We see that, when $\overline{\mathcal{L}}\to0$ the probability distribution
of $\mathcal{L}$ becomes a delta function at zero (the characteristic
function becomes identically one). The distribution function of the
upper bound is\[
\vartheta\left(x\right)\frac{e^{-x/\overline{\mathcal{L}}}}{\overline{\mathcal{L}}}.\]
 Later we will encounter situations where this function gives a good
approximation to the Loschmidt echo probability distribution.

\section{Ising Model in Transverse Field\label{sec:Ising-Model}}

From now on we will give a detailed description of the Loschmidt echo
for the case of an exactly solvable model. The model we consider is
the Ising model in transverse field with Hamiltonian\[
H=-\sum_{i}\left(\sigma_{i}^{x}\sigma_{i+1}^{x}+h\sigma_{i}^{z}\right).\]
 This model can be mapped to quasi-free fermions and so diagonalized
exactly. At zero temperature we distinguish two phases: i) An ordered
one in the longitudinal direction in which $\langle\sigma_{i}^{x}\sigma_{j}^{x}\rangle\stackrel{\left|i-j\right|\to\infty}{\longrightarrow}m^{2}$,
for $\left|h\right|<1$, and ii) A paramagnetic phase for $\left|h\right|>1$.
The points $\left|h\right|=1$ are critical points where the system
is described by a conformal invariant field theory with central charge
$c=1/2$.

The LE is given in this case by \cite{quan06} (superscript, inserted
here for clarity, refer to different values of the coupling constant
$h$)\begin{eqnarray}
\mathcal{L}\left(t\right) & = & \left|\langle\psi^{\left(1\right)}|e^{-itH^{\left(2\right)}}|\psi^{\left(1\right)}\rangle\right|^{2}\nonumber \\
 & = & \prod_{k>0}\left(1-\sin^{2}\left(\vartheta_{k}^{(1)}-\vartheta_{k}^{\left(2\right)}\right)\sin^{2}\left(\Lambda_{k}^{\left(2\right)}t/2\right)\right)\label{eq:LE-ising}\end{eqnarray}
 where $\tan\left(\vartheta_{k}^{\left(i\right)}\right)=-\sin\left(k\right)/\left(h^{\left(i\right)}+\cos\left(k\right)\right)$
and the single particle fermionic dispersion is $\Lambda_{k}^{\left(i\right)}=2\sqrt{\left(h^{\left(i\right)}+\cos\left(k\right)\right)^{2}+\sin\left(k\right)^{2}}$.
The band minimum (maximum) is at $E_{m}=2\min\left\{ \left|1-h^{\left(2\right)}\right|,\left|1+h^{\left(2\right)}\right|\right\} $
($E_{M}=2\max\left\{ \left|1-h^{\left(2\right)}\right|,\left|1+h^{\left(2\right)}\right|\right\} $).
Finally, for periodic boundary conditions that will be used throughout,
the quasi-momenta satisfy $k_{n}=\pi\left(2n+1\right)/L$, $n=0,1,\ldots,L/2-1$.
Exploiting the fact that $H$ decomposes into a direct sum of $L/2$
blocks $4\times4$, we are able to compute the complete dephased equilibrium
state $\rho_{\mathrm{eq}}=\sum_{n}p_{n}|n\rangle\langle n|$. The
result is\begin{equation}
\rho_{\mathrm{eq}}=\sum_{\alpha\in\mathbb{Z}_{2}^{L}}p\left(\alpha\right)|\alpha\rangle\langle\alpha|,\label{eq:rhoeq}\end{equation}
 where the multi-index $\alpha$ is $\alpha=\left(\alpha_{1},\ldots,\alpha_{L}\right)$,
$\alpha_{i}=0,1$, the state is $|\alpha\rangle=\otimes_{k>0}|\alpha_{k}\rangle$
with $|0_{k}\rangle=\cos\left(\vartheta^{\left(2\right)}/2\right)|0,0\rangle_{k,-k}-i\sin\left(\vartheta^{\left(2\right)}/2\right)|1,1\rangle_{k,-k}$
and $|0_{k}\rangle=i\sin\left(\vartheta^{\left(2\right)}/2\right)|0,0\rangle_{k,-k}-\cos\left(\vartheta^{\left(2\right)}/2\right)|1,1\rangle_{k,-k}$.
Finally the weights are given by\begin{equation}
p\left(\alpha\right)=\prod_{k>0}\frac{\tan^{2\alpha_{k}}\left(\delta\vartheta_{k}/2\right)}{1+\tan^{2}\left(\delta\vartheta_{k}/2\right)}.\label{eq:rho_weight}\end{equation}

Using Eq.~(\ref{eq:rhoeq}) together with (\ref{eq:rho_weight})
one can show (cfr.~ref.~\cite{barouch70}) that the dephased state
has the following totally factorized form:\[
\rho_{\mathrm{eq}}=\bigotimes_{k>0}\left(a_{k}|0_{k}\rangle\langle0_{k}|+b_{k}|1_{k}\rangle\langle1_{k}|\right)\]
 where $a_{k}=\left(1+\tan^{2}\left(\delta\vartheta_{k}/2\right)\right)^{-1}$,
and $b_{k}=1-a_{k}$.

\subsection{Short time regime}

Let us first discuss the short-time, transient regime. Looking at
the function $\ln\mathcal{L}\left(t\right)$ one can readily see that
all its $n$-derivatives at $t=0$ are the Riemann sums of a summable
function irrespective of the $h^{\left(i\right)}$s being critical.
This means that all the derivatives of $\ln\mathcal{L}\left(t\right)$
grow linearly with $L$, and together with equation (\ref{eq:losch_cum})
implies that the cumulants are linear even at criticality, i.e.~$\left\langle H^{2n}\right\rangle _{c}\propto L$.
The same result could have been derived by noting that for $\left|h\right|\neq1$
the system is gapful and clustering. When $|\psi^{\left(1\right)}\rangle$
is critical, i.e.~$\left|h^{\left(1\right)}\right|=1$, the scaling
dimension of $\delta H=H^{\left(2\right)}-H^{\left(1\right)}=-\delta h\sum_{i}\sigma_{i}^{z}$
is one and so according to the reasoning in section \ref{sub:Short-time-regime}
the cumulants of $H$ grows linearly with $L$.

Accordingly the CLT applies. Since all the cumulants, including the
variance, grow as $L$, plotting the function $\mathcal{L}_{L}\left(t/\sqrt{L}\right)$
for different sizes $L$, one observes data collapse. This behavior
is illustrated in figure \ref{fig:losch_CLT}. Clearly the plot reproduces
a Gaussian with variance $\lim_{L\to\infty}\left\langle H^{2}\right\rangle _{c}/L$.

\subsection{Long time, large sizes and the order of limits}

\begin{figure}
\noindent \begin{centering}
\includegraphics[width=70mm]{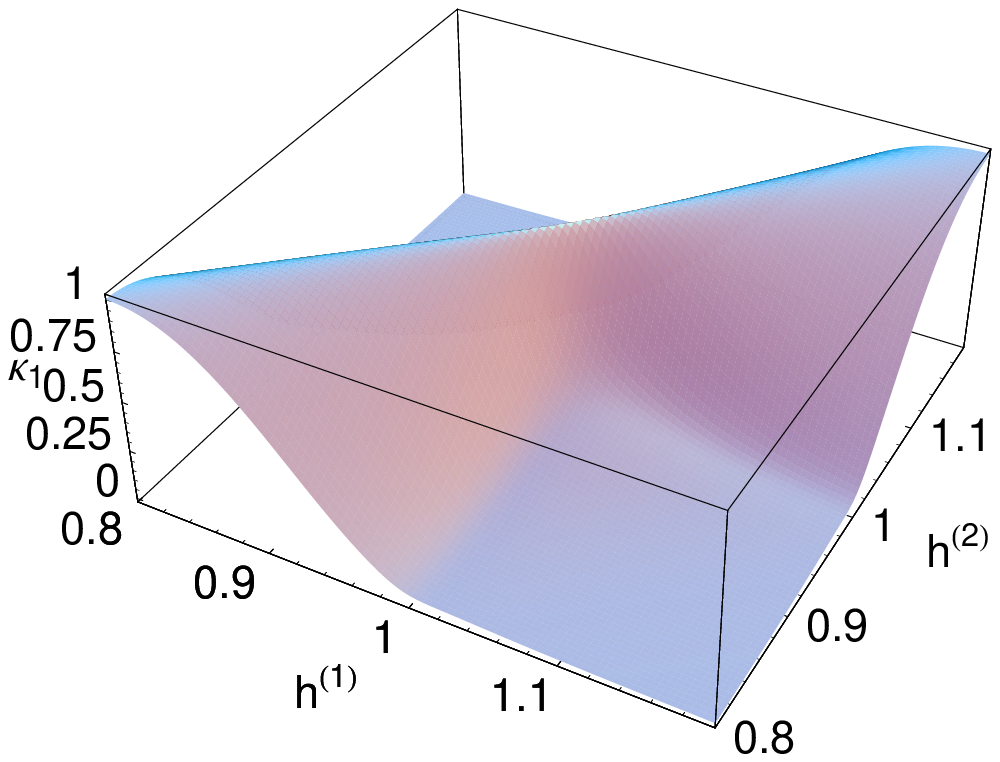} 
\par\end{centering}

\noindent \begin{centering}
\includegraphics[width=70mm]{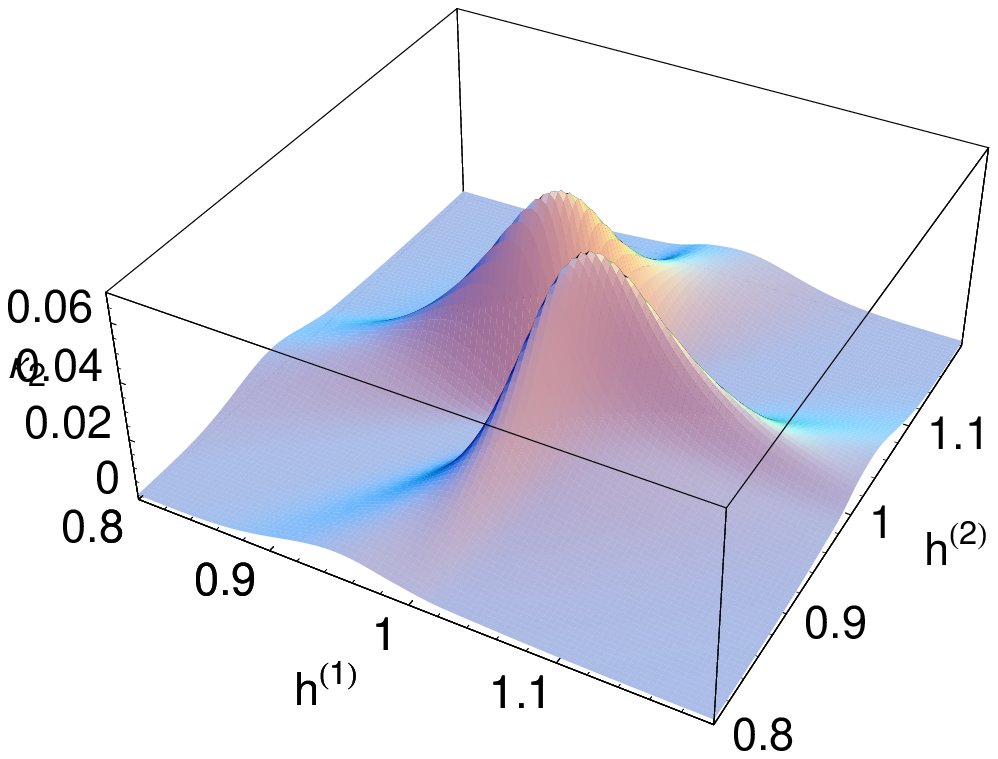} 
\par\end{centering}

\caption{(Color online) From top to bottom, mean and variance of Loschmidt
echo at size $L=100$. The region where the variance is large shrinks
when increasing the system size $L$. The height of the peak however
remains constant. \label{fig:mean-var}}

\end{figure}

Consider now a physical situation where an experimenter computes $\overline{\mathcal{L}_{L}}$.
We inserted the labels $T$ and $L$ to stress that both size and
expectation time are finite, as is required in a true experiment.
Here we want to study the interplay between $T$ and $L$. Consider
first the case where we send $L$ to infinity, or more physically
$L$ is the largest scale of our system. In this situation the spectrum
of $H$ is practically continuous, and we can write the LE as\begin{eqnarray*}
\mathcal{L}\left(t\right) & = & \exp\left[\frac{L}{2\pi}\int_{0}^{\pi}\ln\left(1-\sin^{2}\left(\vartheta_{k}^{(1)}-\vartheta_{k}^{\left(2\right)}\right)\sin^{2}\left(\Lambda_{k}^{\left(2\right)}t/2\right)\right)dk\right].\\
 & \equiv & e^{-Ls\left(t\right)}\end{eqnarray*}
 In this approximation we sent all the revival times to infinity and
so the function $\mathcal{L}\left(t\right)$ is no longer almost periodic,
but rather tends to a precise limit as $t\to\infty$. We can calculate
the limit $s\left(\infty\right)$ and also the first corrections,
as $t\to\infty$. The procedure is outlined in the Appendix. The result
is\begin{equation}
s\left(t\right)=s\left(\infty\right)-\frac{A_{m}}{\left|t\right|^{3/2}}\cos\left(tE_{m}+\frac{3}{4}\pi\right)+\left(m\leftrightarrow M\right),\label{eq:LE-time}\end{equation}
 where $s\left(\infty\right)$ is the limiting value and $A_{m/M}$
are constants which depend on $h^{\left(i\right)}$ and are given
in the appendix. The result (\ref{eq:LE-time_app}) has already been
found in \cite{silva08}, here we provide the explicit form of the
asymptotic value, $s\left(\infty\right)$.

Since the function $\mathcal{L}\left(t\right)=e^{-Ls\left(t\right)}$
has a limit at infinity, its time average is precisely this limit
$\overline{\mathcal{L}}=e^{-Ls\left(\infty\right)}$. More precisely
the distribution function becomes a delta function $P\left(x\right)=\delta\left(x-e^{-Ls\left(\infty\right)}\right)$.

Consider now performing first the time average of Eq.~(\ref{eq:LE-ising}).
According to the discussion in section \ref{sub:Equilibration-and-long},
this requires at least observation times as large as $T\gg L$ (if
we are at criticality). The result in this case is (for the explicit
computation see the following section)\begin{equation}
\overline{\mathcal{L}}=\exp\sum_{k>0}\log\left(1-\sin^{2}\left(\vartheta_{k}^{(2)}-\vartheta_{k}^{\left(1\right)}\right)/2\right)\label{eq:Lbar}\end{equation}
 If now $L$ is large, we can approximate the sum with the integral:
$\overline{\mathcal{L}}=e^{-Lg\left(h^{\left(1\right)},h^{\left(2\right)}\right)}$.
Calling $\delta\vartheta_{k}=\vartheta_{k}^{(2)}-\vartheta_{k}^{\left(1\right)}$
the two functions, $g$ and $s\left(\infty\right)$ are given by \begin{eqnarray}
g & = & -\frac{1}{2\pi}\int_{0}^{\pi}\ln\left(1-\sin^{2}\left(\delta\vartheta_{k}\right)/2\right)dk,\label{eq:g-funz}\\
s\left(\infty\right) & = & -\frac{1}{\pi}\int_{0}^{\pi}\ln\left[\left(1+\left|\cos\left(\delta\vartheta_{k}\right)\right|\right)/2\right]dk.\end{eqnarray}
 We observed that the two averages $e^{-Lg}$ -- obtained by first
doing the time average and then taking large $L$ -- or $e^{-Ls\left(\infty\right)}$
-- obtained by first considering $L$ large and then doing the time
average -- are qualitatively very similar for most values of the parameters
$h^{\left(i\right)}$. The only region where there is an appreciable
difference is when $h^{\left(1\right)}$ and $h^{\left(2\right)}$
correspond to different phases (either $\left|h^{\left(1\right)}\right|<1$
and $\left|h^{\left(2\right)}\right|>1$ or vice-versa).

\subsection{Moments of the Loschmidt echo in presence of degeneracy}

Having the explicit form of the LE Eq.~(\ref{eq:LE-ising}) we can
compute its time average and also other moments. Since the Ising model
is mapped to a free Fermi system on a finite lattice, and given the
form of the quasi-particle dispersion $\Lambda_{k}$, its spectrum
is non-degenerate. In other words the Ising Hamiltonian is 1-non-degenerate.
However, for the same reason, it is not $k$-non-degenerate for $k\ge2$.
This means that to compute moments higher than the first, we really
need to use the explicit form Eq.~(\ref{eq:LE-ising}) and cannot
rely on the results of section \ref{sub:Equilibration-and-long}.
For the first moment this problem does not arise, and we can either
use $\overline{\mathcal{L}}=\sum_{n}p_{n}^{2}$ or do the time average
of Eq.~(\ref{eq:LE-ising}). Correctly the results coincide, and
they rely on the fact that $\sum_{k}\left(n_{k}-m_{k}\right)\Lambda_{k}=0$,
implies $n_{k}=m_{k}$, i.e.~that the spectrum is non-degenerate.
The result is\begin{equation}
\overline{\mathcal{L}}=\prod_{k>0}\left(1-\sin^{2}\left(\vartheta_{k}^{(1)}-\vartheta_{k}^{\left(2\right)}\right)/2\right).\label{eq:LE-mean}\end{equation}
 For later convenience we define $\alpha_{k}=\sin^{2}\left(\vartheta_{k}^{(1)}-\vartheta_{k}^{\left(2\right)}\right)$.
To compute higher moments we first rewrite Eq.~(\ref{eq:LE-ising})
as

\begin{eqnarray*}
\mathcal{L}\left(t\right) & = & \prod_{k>0}\left(1+X_{k}\left(t\right)\right),\\
X_{k}\left(t\right) & = & -\alpha_{k}\sin^{2}\left(\Lambda_{k}t/2\right)=\sum_{\beta=0,\pm1}c_{\beta}^{k}e^{i\beta\Lambda_{k}t}\\
 &  & c_{0}^{k}=-\frac{\alpha_{k}}{2},\quad c_{\pm1}^{k}=\frac{\alpha_{k}}{4}\,.\end{eqnarray*}
 We write the $n$-th power of the LE as\begin{eqnarray*}
\left[\mathcal{L}\left(t\right)\right]^{n} & = & \prod_{k>0}\left(1+Y_{k}^{\left(n\right)}\left(t\right)\right),\\
Y_{k}^{\left(n\right)}\left(t\right) & = & \sum_{m=1}^{n}\left(\begin{array}{c}
n\\
m\end{array}\right)\left[X_{k}\left(t\right)\right]^{m}\equiv\sum_{\gamma=0,\pm1,\cdots,\pm m}g_{\gamma,k}^{\left(n\right)}e^{i\gamma\Lambda_{k}t}.\end{eqnarray*}
 Now, when computing $M$ products of $Y_{k}^{\left(n\right)}$ terms,
only the $\gamma=0$ term will survive after taking the time average.
In other words: \[
\overline{Y_{k_{1}}^{\left(n\right)}\cdots Y_{k_{M}}^{\left(n\right)}}=g_{0,k_{1}}^{\left(n\right)}\cdots g_{0,k_{M}}^{\left(n\right)}\]
 and so we have\[
\overline{\left[\mathcal{L}\left(t\right)\right]^{n}}=\prod_{k>0}\left(1+g_{0,k}^{\left(n\right)}\right).\]

An explicit formula for $g_{0,k}^{\left(n\right)}$ is \[
g_{0,k}^{\left(n\right)}=\sum_{m=1}^{n}\left(\begin{array}{c}
n\\
m\end{array}\right)\sum_{\stackrel{\beta_{1},\ldots\beta_{m}}{\sum\beta_{i}=0}}c_{\beta_{1}}^{k}\cdots c_{\beta_{m}}^{k}.\]
 Noting that\[
\sum_{\stackrel{\beta_{1},\ldots\beta_{m}}{\sum\beta_{i}=0}}c_{\beta_{1}}^{k}\cdots c_{\beta_{m}}^{k}=\sum_{\substack{n_{1}\\
n_{0}+2n_{1}=m}
}\frac{m!}{\left(n_{1}!\right)^{2}n_{0}!}\left(c_{0}^{k}\right)^{n_{0}}\left(c_{1}^{k}\right)^{2n_{1}},\]
 we obtain\begin{eqnarray*}
g_{0,k}^{\left(n\right)} & = & \sum_{m=1}^{n}\left(\frac{-\alpha_{k}}{4}\right)^{m}\left(\begin{array}{c}
n\\
m\end{array}\right)\frac{\left.\partial_{t}^{m}\left(2t-t^{2}-1\right)^{m}\right|_{t=0}}{m!}\\
 & = & \sum_{m=1}^{n}\left(\frac{-\alpha_{k}}{4}\right)^{m}\left(\begin{array}{c}
n\\
m\end{array}\right)\left(\begin{array}{c}
2m\\
m\end{array}\right).\end{eqnarray*}
 For example we have\begin{eqnarray*}
g_{0,k}^{\left(1\right)} & = & -\frac{\alpha_{k}}{2}\\
g_{0,k}^{\left(2\right)} & = & -\alpha_{k}+\frac{3}{8}\alpha_{k}^{2}\\
g_{0,k}^{\left(3\right)} & = & -\frac{3}{2}\alpha_{k}+\frac{9}{8}\alpha_{k}^{2}-\frac{5}{16}\alpha_{k}^{3}\\
g_{0,k}^{\left(4\right)} & = & -2\alpha_{k}+\frac{9}{4}\alpha_{k}^{2}-\frac{5}{4}\alpha_{k}^{3}+\frac{35}{128}\alpha_{k}^{4}.\end{eqnarray*}
 The variance of the LE is then given by\begin{equation}
\overline{\Delta\mathcal{L}^{2}}=\prod_{k>0}\left(1-\alpha_{k}+\frac{3}{8}\alpha_{k}^{2}\right)-\prod_{k>0}\left(1-\alpha_{k}+\frac{1}{4}\alpha_{k}^{2}\right).\label{eq:LE-var}\end{equation}

The first moment and the variance are plotted in figure \ref{fig:mean-var}.
One should note that close to the critical points $h_{c}=\pm1$ there
appears a small region $\delta h$ where the variance is large.

Equation (\ref{eq:LE-var}) gives explicitly the variance in a case
where the non-resonant hypothesis is violated. Since $\left|\alpha_{k}\right|\le1$
generally the variance is given by the difference between two exponentially
small quantities and so, a fortiori, is exponentially small in the
system size $L$. However, looking at figure \ref{fig:mean-var} one
notes a small region of parameter close to the critical points, where
the variance is large. As we will see this fact has important consequences. 

It is now interesting to compare the result in equation (\ref{eq:LE-var})
with what would have been obtained assuming a non-resonant condition.
Clearly the second moment computed assuming non resonance has less
terms than the correct one. Since for the LE all the contributions
are positive, the non-resonant result ought to be smaller. In other
words, for the variance we must have $\overline{\Delta\mathcal{L}^{2}}\ge\overline{\Delta\mathcal{L}_{\mathrm{nr}}^{2}}$.

\paragraph*{Comparison with the non-resonant result}

To compute the variance assuming non-resonance we use \begin{eqnarray*}
\overline{\mathcal{L}_{\mathrm{nr}}^{2}} & = & \overline{\mathcal{L}}^{2}+2\sum_{i<j}p_{i}^{2}p_{j}^{2}\\
 & = & 2\overline{\mathcal{L}}^{2}-\sum_{i}p_{i}^{4}\end{eqnarray*}
 Using this formula together with Eq.~(\ref{eq:rho_weight}) we obtain\begin{eqnarray*}
\overline{\Delta\mathcal{L}_{\mathrm{nr}}^{2}} & = & \overline{\mathcal{L}}^{2}-\prod_{k>0}\left(1-\alpha_{k}+\frac{1}{8}\alpha_{k}^{2}\right)\\
 & = & \prod_{k>0}\left(1-\alpha_{k}+\frac{1}{4}\alpha_{k}^{2}\right)-\prod_{k>0}\left(1-\alpha_{k}+\frac{1}{8}\alpha_{k}^{2}\right).\end{eqnarray*}

We have verified that the inequality $\overline{\Delta\mathcal{L}^{2}}\ge\overline{\Delta\mathcal{L}_{\mathrm{nr}}^{2}}$
holds for all values of the coupling constants $h^{\left(1\right)}$
and $h^{\left(2\right)}$. However the qualitative behavior of $\overline{\Delta\mathcal{L}_{\mathrm{nr}}^{2}}$
is very similar to the true variance $\overline{\Delta\mathcal{L}^{2}}$.

\subsection{The Loschmidt echo distribution function}

We now turn to consider the whole probability distribution of the
LE. As we have noted earlier, for any $L$ finite being the spectrum
discrete, $\mathcal{L}\left(t\right)$ is an almost periodic function.
Actually $\mathcal{L}\left(t\right)$ belongs to a smaller class,
since it is a trigonometric polynomial. In any case, most results
we will present are valid for the larger class of almost periodic
functions.

\begin{figure}
\noindent \begin{centering}
\includegraphics[width=7cm]{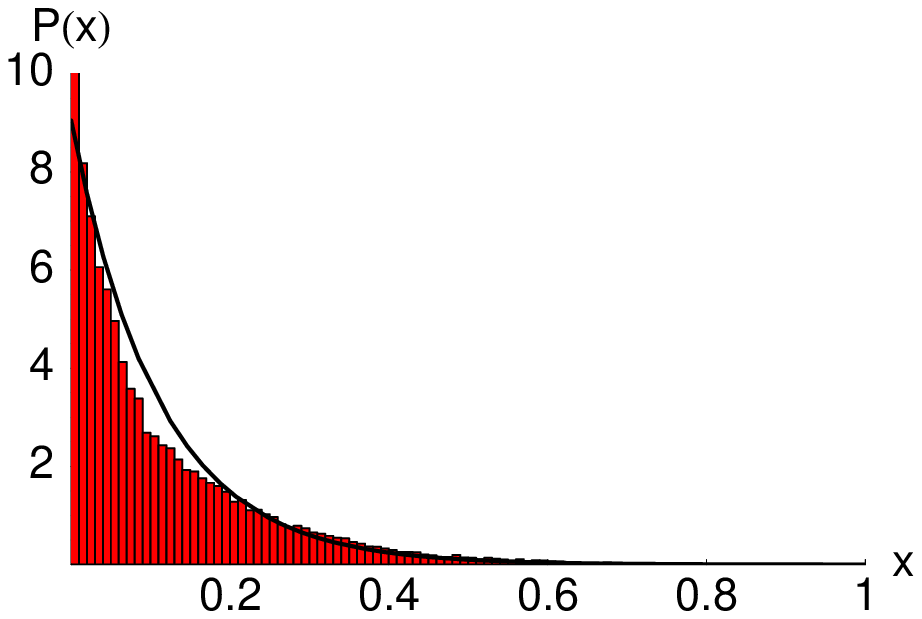}
\par\end{centering}

\begin{centering}
\includegraphics[width=7cm]{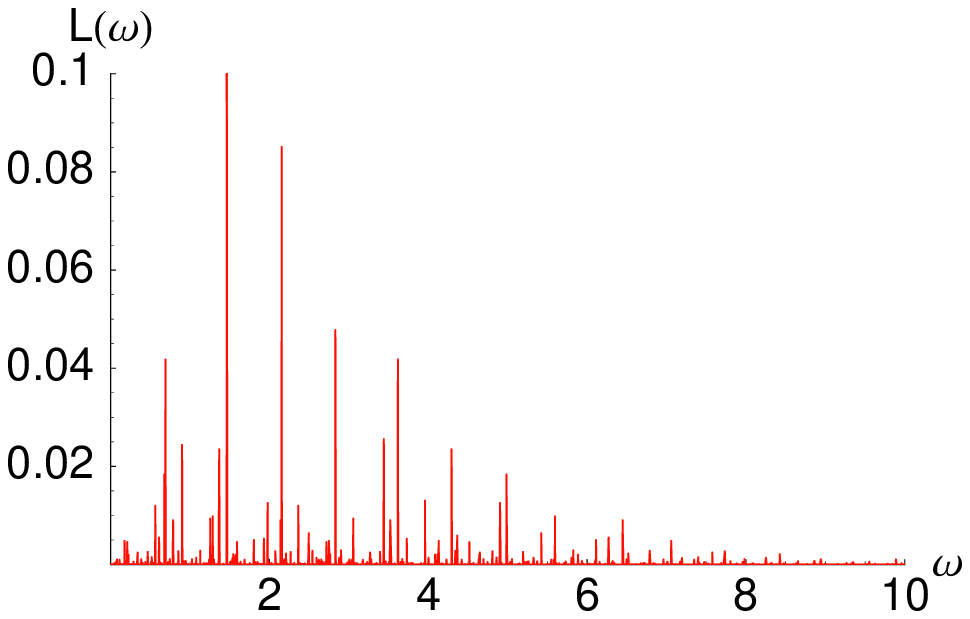}
\par\end{centering}

\caption{(Color online) Approximate exponential behavior. Parameters are $L=18,\, h^{(1)}=0.3,\, h^{(2)}=1.4$.
When $\delta h$ is large this behavior is observed even for moderate
sizes (here $L=18$), (upper panel). The thick line reproduces $\vartheta\left(x\right)e^{-x/\overline{\mathcal{L}}}/\overline{\mathcal{L}}$
with $\overline{\mathcal{L}}$ given by equation (\ref{eq:LE-mean}).
In the lower panel we plot the Fourier series $\hat{\mathcal{L}}_{\mathrm{disc}}\left(\omega\right)$
given by Eq.~(\ref{eq:L-omega}). \label{fig:exponential}}

\end{figure}

\begin{figure}
\noindent \begin{centering}
\includegraphics[width=7cm]{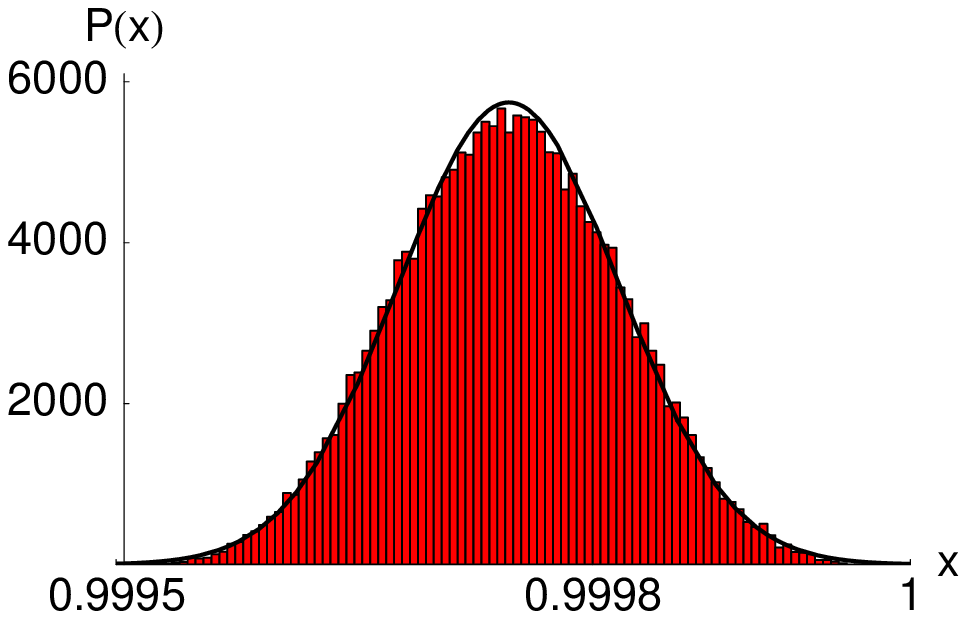} 
\par\end{centering}

\noindent \begin{centering}
\includegraphics[width=7cm]{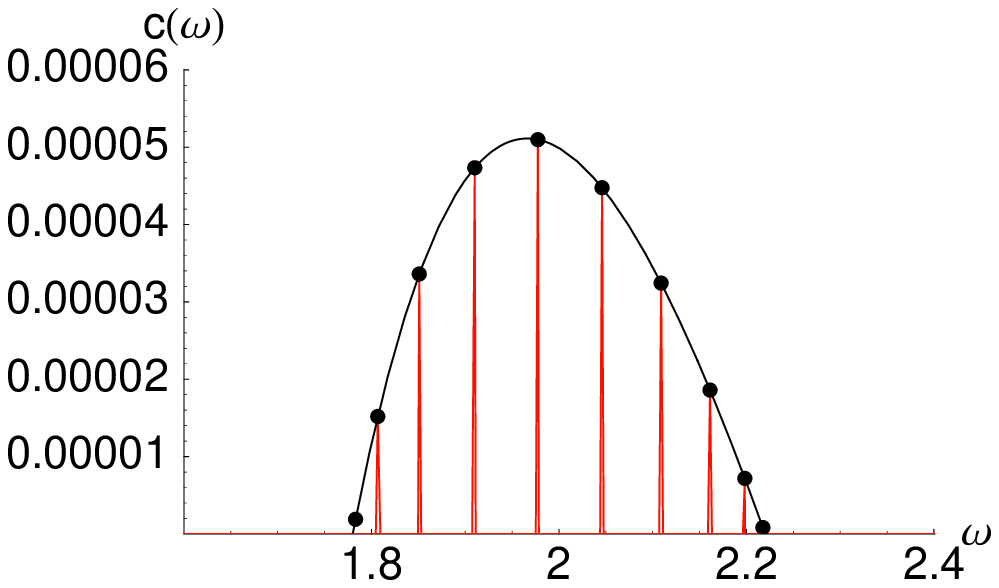} 
\par\end{centering}

\caption{(Color online) Gaussian behavior for $\delta h$ small but away from
criticality. Parameters are $L=20,\, h^{(1)}=0.1,\, h^{(2)}=0.11$.
Note that the distribution function is an extremely peaked Gaussian
(upper panel). The thick line is a Gaussian with mean and variance
given by Eqns.~(\ref{eq:LE-mean}) and (\ref{eq:LE-var}). In the
bottom panel one can notice that many frequencies contribute to the
LE. The one particle contribution $c\left(\omega\right)$, Eq.~(\ref{eq:c-omega}),
is given by the black dots while the red curve gives the true spectral
decomposition $\hat{\mathcal{L}}_{\mathrm{disc}}\left(\omega\right)$
given by Eq.~(\ref{eq:L-omega}). \label{fig:gaussian}}

\end{figure}

\begin{figure}
\noindent \begin{centering}
\includegraphics[width=7cm]{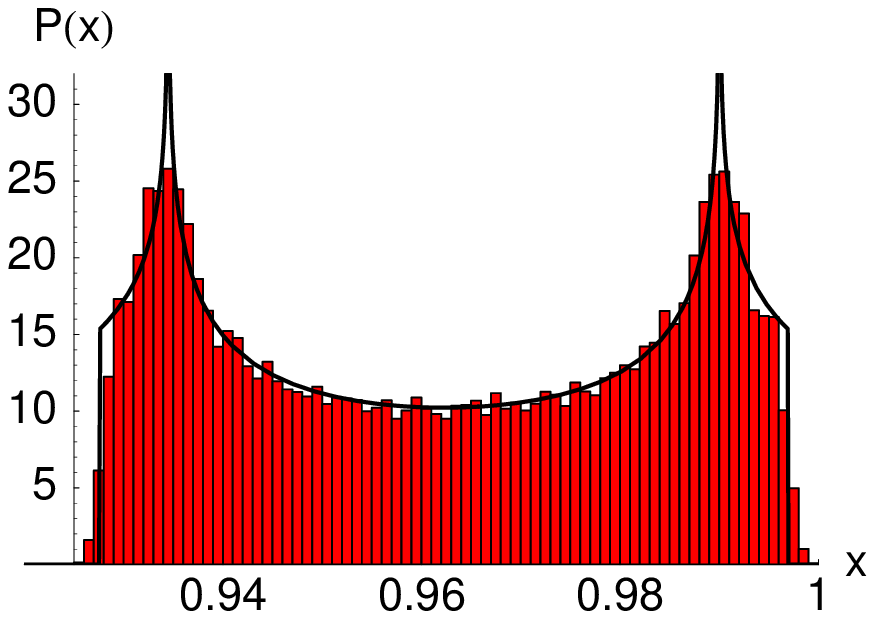} 
\par\end{centering}

\noindent \begin{centering}
\includegraphics[width=7cm]{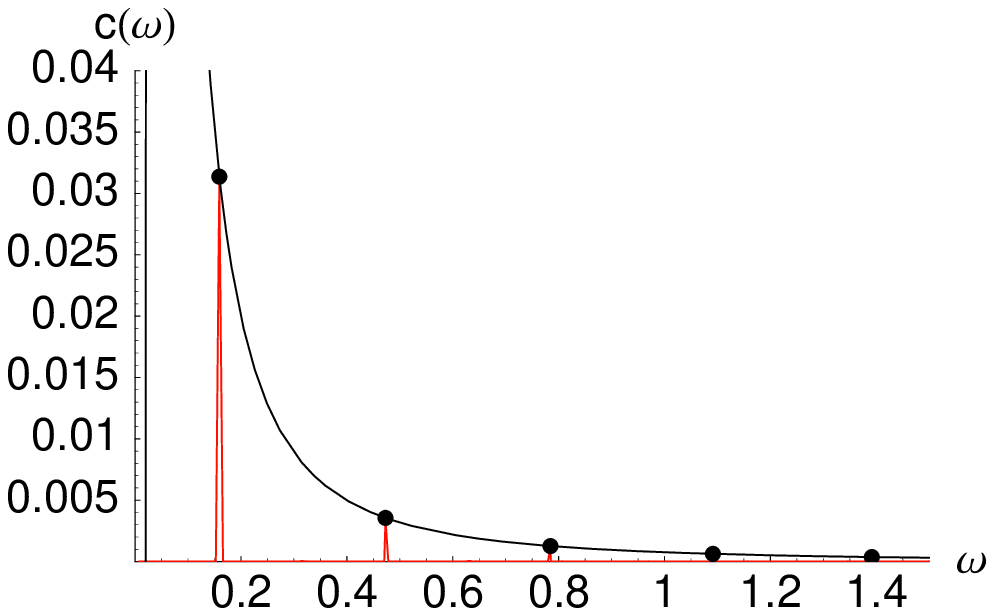} 
\par\end{centering}

\caption{(Color online) Typical {}``batman-hood'' behavior. Parameters are
$L=40,\, h^{(1)}=0.99,\, h^{(2)}=1.01$. Upper panel: probability
distribution (histogram) together with the result of the approximation
given in Eq.~\ref{eq:LE-two-freq} (thick line). The mean $\overline{\mathcal{L}}$
is taken from Eq.~(\ref{eq:LE-ising}) while the parameters $A$
and $B$ are obtained by computing the two largest spectral weights
(see equation (\ref{eq:LE-app})). Lower panel: the red curve is the
Fourier series $\hat{\mathcal{L}}_{\mathrm{disc}}\left(\omega\right)$
(projected spectral density), black curve shows the highest coefficient
$c\left(\omega\right)$ given by Eq.~(\ref{eq:c-omega}), together
with allowed frequencies (black dots). \label{fig:double-peak}}

\end{figure}

\begin{figure}
\noindent \begin{centering}
\includegraphics[width=7cm]{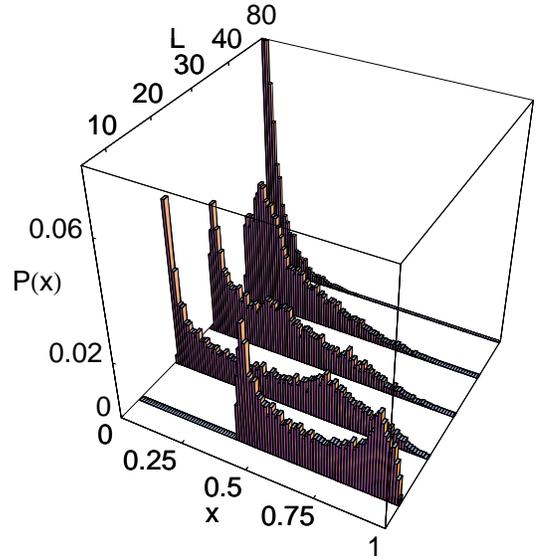} 
\par\end{centering}

\caption{(Color online) Batman-hood distribution approaching an exponential
one when increasing system size $L$ at fixed $h^{\left(i\right)}$.
Parameters are $h^{(1)}=0.9,\, h^{(2)}=1.2$ and chain length are
$L=10,20,30,40$.\label{fig:hat-exp}}

\end{figure}

\begin{figure}
\noindent \begin{centering}
\includegraphics[width=7cm]{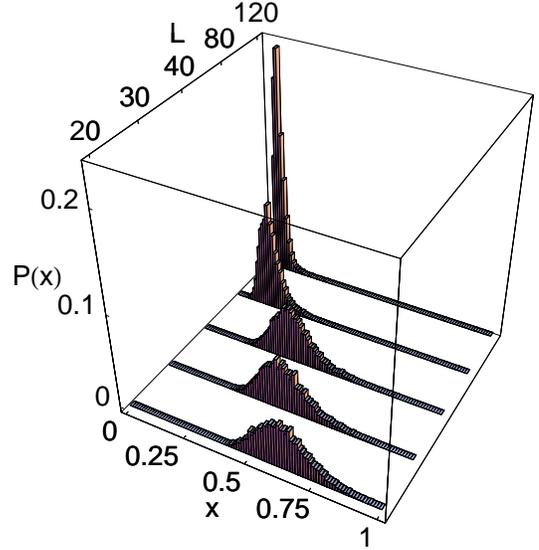} 
\par\end{centering}

\caption{(Color online) Gaussian distribution approaching an exponential one
when increasing system size $L$ at fixed $h^{\left(i\right)}$. Parameters
are $h^{(1)}=0.2,\, h^{(2)}=0.6$ and chain length are $L=20,30,40,80,120$.\label{fig:gauss-exp}}

\end{figure}

We now give the results for the whole LE probability distribution
function. We have observed three kinds of universal behavior emerging
in different, well defined regimes. i) Exponential behavior where
the probability distribution is well approximated by $\vartheta\left(x\right)e^{-x/\overline{\mathcal{L}}}/\overline{\mathcal{L}}$
(figure \ref{fig:exponential}), ii) Gaussian behavior, (figure \ref{fig:gaussian}),
and iii) A universal double peaked, {}``batman-hood'' shaped function,
(figure \ref{fig:double-peak}). More precisely we have the following
scenario: 
\begin{itemize}
\item $\delta h$ large. In this case, for $L$ moderately large, the distribution
is approximately exponential. The feature is more pronounced when
$h^{\left(1\right)},\, h^{\left(2\right)}$ are in different phases,
the limiting case being $h^{\left(1\right)}\approx\pm h^{\left(2\right)}$. 
\item $\delta h$ small. In this case we have to distinguish two situations:

\begin{itemize}
\item $h^{\left(i\right)}$ close to the critical point:

\begin{itemize}
\item $L\ll\left|h^{\left(i\right)}-1\right|^{-1}\propto\xi$, universal
batman-hood distribution. Note that $L\ll\xi$ is the so called quasi-critical
regime. 
\item $L\gg\left|\delta h\right|^{-1}$, exponential distribution 
\end{itemize}
\item Off critical:

\begin{itemize}
\item $L\gg\left|\delta h\right|^{-2}$, exponential distribution 
\item Otherwise Gaussian. 
\end{itemize}
\end{itemize}
\end{itemize}
In other words, say that we fixed $h^{\left(i\right)}$ in order to
have either a Gaussian or a double-peaked distribution. We can always
find an $L$ large enough such that the distribution becomes exponential
in both cases. However, for the Gaussian case we must reach considerably
larger sizes $L\gg\left|\delta h\right|^{-2}$ (compare figures \ref{fig:hat-exp}
and \ref{fig:gauss-exp}).

We have observed an exponential distribution in the region of parameters
where the average LE is much smaller than one: $\overline{\mathcal{L}}\ll1$.
Due to the bound $\overline{\mathcal{L}^{2}}<2\overline{\mathcal{L}}^{2}$,
one has $\Delta\mathcal{L}<\overline{\mathcal{L}}^{2}$ so that when
$\overline{\mathcal{L}}\ll1$ even the variance is small. Since the
LE is supported in $\left[0,1\right]$ and in particular $\mathcal{L}\left(t\right)$
must be positive we expect in the region $\overline{\mathcal{L}}\ll1$
a distribution with positive support, with a large peak very close
to zero, and rapidly decaying tail. We have verified that an exponential
distribution of the form $\vartheta\left(x\right)e^{-x/\overline{\mathcal{L}}}/\overline{\mathcal{L}}$
gives a pretty good approximation in the region $\overline{\mathcal{L}}\ll1$.
Note in any case, that the exponential form is always an approximation.
In particular, for $x\to0$ the true distribution $P\left(x\right)$
tends to zero for any value of the parameters. This happens since
we have always $\mathcal{L}\left(t\right)>0$ strictly. And so generally
$0<\mathcal{L}_{\mathrm{min}}\le\mathcal{L}\le1$. This feature can
be accounted for by adding a (small) $\epsilon$ term to the exponential:
$P\left(x\right)\propto e^{-x/\overline{\mathcal{L}}-\epsilon/x}$.

Let us now investigate the conditions under which the first moment
is small and so we expect an approximately exponential behavior. Looking
at equation (\ref{eq:Lbar}) we see that $\overline{\mathcal{L}}\ll1$
holds when $Lg\left(h^{\left(1\right)},h^{\left(2\right)}\right)\gg1$
where $g$ is given by Eq.~(\ref{eq:g-funz}). Clearly the function
$g$ is zero (its minimum) when $h^{\left(1\right)}=h^{\left(2\right)}$,
and is quadratic in the difference on the diagonal. Quite interestingly
the function $g$ is appreciably different from zero only when $h^{\left(1\right)}$
and $h^{\left(2\right)}$ correspond to different phases (i.e.~either
$\left|h^{\left(1\right)}\right|<1$ and $\left|h^{\left(2\right)}\right|>1$
or vice-versa) so that in these cases we observe exponential behavior
even for moderately small lattices. When $h^{\left(1\right)}$ is
close to $h^{\left(2\right)}$, but away from critical points, the
function $g$ is quadratic in the difference $\delta h$ so that $\overline{\mathcal{L}}\approx\exp\left(-\mathrm{const.}\times L\delta h^{2}\right)$.
Hence to have $\overline{\mathcal{L}}\ll1$ and so to observe approximate
exponential behavior we obtain the relation $L\gg\delta h^{-2}$.
At criticality and for $\delta h$ small instead, we can use $\overline{\mathcal{L}}\approx F^{4}$
where $F$ is the fidelity which scales as (see section \ref{sub:Short-time-regime}
and ref.~\cite{LCV07}) . In the quantum Ising model we are considering
we have $d=\zeta=\nu=\Delta=1$ and so the average behaves as $\overline{\mathcal{L}}\approx\exp\left(-\mathrm{const.}\times L^{2}\delta h^{2}\right)$.
This means that for $\delta h$ small around a critical point $h_{c}=\pm1$,
the condition to have an exponential distribution becomes $L\gg\delta h^{-1}.$ 

We now turn to consider the origin of the batman-hood shaped distribution
function. As we have already noticed, the LE is a (finite) sum of
cosines with given frequencies and amplitude. We can imagine a situation
where only few frequencies contribute to the LE. In the limiting case,
only two non-zero terms. That means that the LE can be approximated
by\begin{equation}
\mathcal{L}\left(t\right)=\overline{\mathcal{L}}+A\cos\left(\omega_{A}t\right)+B\cos\left(\omega_{B}t\right).\label{eq:LE-two-freq}\end{equation}
where we can assume $A,\, B$ positive.

We have devoted some attention to the probability distribution generated
by such a function. If $\omega_{A}$ and $\omega_{B}$ are rationally
dependent, the function is periodic and the distribution function
has square root singularities at all values of $\mathcal{L}$ where
$\partial_{t}\mathcal{L}\left(t\right)=0$. However in our case the
frequencies $\omega_{A/B}$ are always rationally independent. In
this case the vector $\boldsymbol{x}\left(t\right)=\left(\omega_{A}t,\omega_{B}t\right)$
wraps around the torus in a uniform way. We can then invoke ergodicity
and transform the time average into a {}``phase space'' average
(in this case the phase space is $\boldsymbol{x}=\left(x_{1},x_{2}\right)$).
Hence the probability distribution function is given by\begin{align*}
P\left(\mathcal{L}\left(t\right)=\chi\right) & =\frac{1}{\left(2\pi\right)^{2}}\int_{0}^{2\pi}dx_{1}\int_{0}^{2\pi}dx_{2}\delta\left(\mathcal{L}\left(x_{1},x_{2}\right)-\chi\right).\end{align*}
By using Eq.~(\ref{eq:LE-two-freq}) this probability density can
be written as \begin{equation}
P\left(\chi+\overline{\mathcal{L}}\right)=\frac{1}{\pi^{2}A}\int_{\max\left\{ -1,\frac{\left(\chi-B\right)}{A}\right\} }^{\min\left\{ 1,\frac{\left(\chi+B\right)}{A}\right\} }\frac{dz}{\sqrt{\left(\frac{\chi+B}{A}-z\right)\left(z-\frac{\chi-B}{A}\right)\left(1-z^{2}\right)}}.\label{eq:bat}\end{equation}
 The integral above can be expressed in terms of elliptic functions,
but we won't need its explicit expression. Typically the function
(\ref{eq:bat}) is batman-hood shaped function (see fig.~\ref{fig:double-peak}),
with support in $\left[\overline{\mathcal{L}}-\left|A+B\right|,\overline{\mathcal{L}}+\left|A+B\right|\right]$
and two peaks at $\chi=\overline{\mathcal{L}}\pm\left|A-B\right|$.
The divergence at the peaks position is of logarithmic type, close
to the peaks (i.e.~$\chi=\overline{\mathcal{L}}\pm\left|A-B\right|+\epsilon$)
one has\[
P\left(\chi\right)=-\frac{\ln\left(\epsilon\right)}{2\pi^{2}\sqrt{\left|AB\right|}}+O\left(1\right).\]
 Note that we never observe the limiting case where $B=0$ and $\mathcal{L}\left(t\right)$
becomes a periodic function. This means that even a very small spectral
weight on $B$ cannot be discarded. On the other hand the distribution
function (\ref{eq:bat}) seems to be quite stable against the presence
of other oscillating terms with small spectral weight. This stability
can be seen in fig.~\ref{fig:double-peak} where one clearly has
at least three frequencies with reasonable spectral weight, but the
probability density is still well approximated by a batman-hood.

\paragraph*{Spectral analysis\label{sub:Spectral-analysis}}

To understand the behavior of $P\left(\mathcal{L}=x\right)$ we do
a spectral analysis of $\mathcal{L}\left(t\right)$ to see which frequencies
contribute most. In fact, for almost periodic functions there is a
similar Fourier decomposition as for periodic functions. The Fourier
expansion is given in this case by $\hat{\mathcal{L}}_{\mathrm{disc}}\left(\omega\right)=\overline{\mathcal{L}\left(t\right)e^{i\omega t}}$.
Taking into account Eq.~(\ref{eq:LE-base}) $\hat{\mathcal{L}}_{\mathrm{disc}}\left(\omega\right)$
can be written as\begin{equation}
\hat{\mathcal{L}}_{\mathrm{disc}}\left(\omega\right)=\sum_{n,m}p_{n}p_{m}\delta_{\omega,E_{n}-E_{m}}\label{eq:L-omega}\end{equation}

We would like to know which frequencies have the largest weight. This
is achieved by expanding the product in Eq.~(\ref{eq:LE-ising})
\footnote{A more precise approximation valid also in region where $\delta h$
is large, is given by $\mathcal{L}\left(t\right)=\prod_{k>0}\left(1+c_{0}^{k}+\tilde{X}_{k}\left(t\right)\right)\simeq\overline{\mathcal{L}}\left(1+\sum_{k>0}\left(1+c_{0}^{k}\right)^{-1}\tilde{X}_{k}\left(t\right)\right)$.
The two expressions coincide when $\delta h$ is small. %
} \begin{eqnarray*}
\mathcal{L}\left(t\right) & = & 1+\sum_{k>0}X_{k}\left(t\right)+\sum_{k_{1},k_{2}>0}X_{k_{1}}\left(t\right)X_{k_{2}}\left(t\right)+\cdots\\
 & = & \overline{\mathcal{L}}+\sum_{k>0}\tilde{X}_{k}\left(t\right)+\sum_{k_{1},k_{2}>0}\tilde{X}_{k_{1}}\left(t\right)\tilde{X}_{k_{2}}\left(t\right)+\cdots\\
\mathrm{with} &  & \tilde{X}_{k}\left(t\right)=\sum_{\beta=\pm1}c_{\beta}^{k}e^{i\beta\Lambda_{k}t}=\frac{\alpha_{k}}{2}\cos\left(\Lambda_{k}t\right).\end{eqnarray*}
 Now, since each $\tilde{X}_{k}\left(t\right)$ is smaller than $1/2$
in modulus, it is reasonable to approximate the LE with the first
two terms of this expansion and we obtain\begin{equation}
\mathcal{L}\left(t\right)\simeq\overline{\mathcal{L}}+\sum_{k>0}\frac{\alpha_{k}}{2}\cos\left(\Lambda_{k}t\right).\label{eq:LE-app}\end{equation}
 In this approximation we only wrote the zero-frequency contribution,
which corresponds to the mean, and the contribution coming from the
one particle spectrum. The next term has also contributions coming
from the two particle spectrum. To be more precise, call $E^{\left(n\right)}$
the energy of the $n$-particle spectrum then $E_{a}^{\left(1\right)}-E_{b}^{\left(0\right)}\propto\Lambda_{k}$,
(first order contribution), while $E_{a}^{\left(1\right)}-E_{b}^{\left(1\right)}\propto\Lambda_{k_{1}}-\Lambda_{k_{2}}$,
and $E_{a}^{\left(2\right)}-E_{b}^{\left(0\right)}\propto\Lambda_{k_{1}}+\Lambda_{k_{2}}$(second
order contribution with less spectral weight).

Note that we expect Eq.~(\ref{eq:LE-app}) to be approximately valid
(with a different form for the amplitudes and the frequencies) also
for non integrable models in which a one-particle approximation works
well.

Now, if there is a regime where the amplitudes $\alpha_{k}/2$ are
highly peaked around few quasi-momenta, in the limiting case only
two, then the LE can be approximated as in Eq.~(\ref{eq:LE-two-freq})
and we expect a double peaked distribution function. So we are led
to study the (one-particle) amplitude function\begin{equation}
c\left(\omega\right)\equiv\left.\frac{\alpha_{k}}{2}\right|_{\omega=\Lambda_{k}},\,\,\omega\in\left[E_{m},E_{M}\right].\label{eq:c-omega}\end{equation}
 Generally $c\left(\omega\right)$ is a bell-shaped function, starting
linearly from the band minimum $E_{m}$, reaching a maximum value
and then decreasing to zero at the band maximum $E_{M}$. It is not
difficult to show that \cite{nota}, when $\delta h$ is small and
for roughly $\left|1-h^{(2)}\right|\lesssim10^{-1}$, $c\left(\omega\right)$
starts developing a peak, the width of which being proportional to
$\left|1-h^{(2)}\right|$. In the limiting case $h^{(2)}=1$, $c\left(\omega\right)$
has its maximum at $E_{m}=0$ and then decreases monotonically to
$E_{M}=2$. So, for $\delta h$ small and for $\left|1-h^{(2)}\right|\lesssim10^{-1}$,
$c\left(\omega\right)$ is a peaked function. In order to have few
frequencies fall within the peak, and so to have large spectral weight
on few frequencies, we must additionally have $L\left|1-h^{(2)}\right|\ll1$.
This is easily seen analyzing the dispersion $\Lambda_{k}$ for $h^{(2)}$
close to the critical point \cite{nota-lambda}. All in all, the conditions
to have a batman-hood distribution, are $\delta h$ small and $L\left|1-h^{(2)}\right|\ll1$.
The feature is more pronounced when the $h_{i}$ are not precisely
critical. In fact even though $c\left(\omega\right)$ is most peaked
when $h^{(2)}=1$ (and the peak is at $\omega=0$), we have to remember
that the allowed values of $\omega$ are $\omega_{n}=\Lambda_{k_{n}}$
where $k_{n}=\pi\left(2n+1\right)/L$, and the smallest frequency
is $\omega_{1}=\Lambda_{\pi/L}$. If we perturb $h^{(2)}$ from 1
the peak of $c\left(\omega\right)$ shifts to the right, approaching
$\omega_{1}$, so it is favorable to have $h^{(2)}\neq1$. Not surprisingly,
the conditions to have a batman-hood probability density, coincide
with having a large variance (see figure \ref{fig:mean-var} bottom
panel).

Generally fixing $h^{\left(i\right)}$ and increasing the size $L$,
one eventually violates the quasi-critical condition $L\ll\xi$. At
this stage the double-peak feature tends to disappear and the distribution
approaches an exponential one. This can be clearly seen in figure
\ref{fig:hat-exp}. From this figure one can have the impression that
the {}``double-peak feature'' is a prerequisite of short sizes,
since in this case one has few frequencies anyway. As we have tried
to explain instead, this feature survives for larger sizes, provided
we shrink $\delta h$ sufficiently (figure \ref{fig:double-peak}).

Instead, when $\delta h$ is small but $h^{\left(i\right)}$ are far
from the critical point, then $c\left(\omega\right)$ is not peaked,
and many frequencies have a large spectral weight (see figure \ref{fig:gaussian}).
In this case the distribution becomes Gaussian.The emergence of a
Gaussian distribution can be qualitatively understood in the following
way. First write the LE according to its spectral decomposition $\mathcal{L}\left(t\right)=\sum_{n}A_{n}e^{it\omega_{n}}$
where the amplitudes are precisely given by $A_{n}=\hat{\mathcal{L}}_{\mathrm{disc}}\left(\omega_{n}\right)$
and are positive. When the frequencies are rationally independent
the variables $x_{n}=t\omega_{n}$ wrap uniformly around a large dimensional
torus. Then one can consider each $A_{n}\times e^{tx_{n}}$ as an
independent random variable. The assumption $\delta h$ is small but
$h^{\left(i\right)}$ away from criticality corresponds to say that
$\mathcal{L}\left(t\right)$ can be considered as a sum of \emph{many}
independent random variables, giving rise to a Gaussian distribution.
When $L\gg\left|\delta h\right|^{-2}$ the conditions of independence
breaks down and we recover an approximate exponential behavior. 

\section{Probability distribution function for the magnetization}

In the same spirit we can compute the probability distribution of
a local operator. The first candidate that comes to mind is the transverse
magnetization. We computed the following time dependent observable
$m\left(t\right)=\left\langle \psi^{\left(1\right)}|e^{itH^{\left(2\right)}}\sigma_{i}^{z}e^{-itH^{\left(2\right)}}|\psi^{\left(1\right)}\right\rangle $.
Using again equation (\ref{eq:rhoeq}) one obtains %
\footnote{This is precisely Eq.~(5.6) of \cite{barouch70} in the zero temperature
limit. %
}

\begin{equation}
m\left(t\right)=\frac{1}{L}\sum_{k}\cos\left(\vartheta_{k}^{\left(2\right)}\right)\cos\left(\delta\vartheta_{k}\right)+\sin\left(\vartheta_{k}^{\left(2\right)}\right)\sin\left(\delta\vartheta_{k}\right)\cos\left(t\Lambda_{k}^{\left(2\right)}\right),\label{eq:mt}\end{equation}
 where the quasi-momenta range now in the whole Brillouine zone: $k=\pi\left(2n+1\right)/L,$
$n=0,1,\ldots,L-1$. Correctly, when $h^{\left(1\right)}=h^{\left(2\right)}$
we recover the zero temperature equilibrium result $\langle\sigma_{i}^{z}\rangle=L^{-1}\sum_{k}\cos\left(\vartheta_{k}\right)$.

\begin{figure}
\noindent \begin{centering}
\includegraphics[width=7cm]{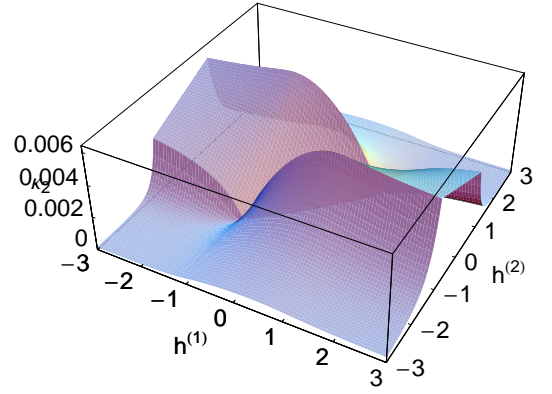} 
\par\end{centering}

\caption{(Color online) Variance of the magnetization as given by Eq.~(\ref{fig:varm})
for $L=80$. A signature of criticality are the cusps at $h^{\left(2\right)}=\pm1$.
\label{fig:varm}}

\end{figure}

From equation (\ref{eq:mt}) we see that $m\left(t\right)$ can be
written ---exactly--- as a constant term plus an an oscillating part
with frequencies given by the single particle spectrum $\Lambda_{k}$.
The discussion on characteristic times becomes simplified as all time
scales are uniquely determined by $\Lambda_{k}^{\left(2\right)}$.
For example the time $T_{1}$ necessary to observe the correct mean:
$\overline{m^{T_{1}}}=\overline{m}$ must simply satisfy $T_{1}\gg\mathrm{gap}^{-1}$
which means $T_{1}\gg L$ in the quasi-critical regime $\left|h^{\left(2\right)}-1\right|^{-1}\gg L$,
while it suffices to have $T_{1}\gg O\left(1\right)$ away from criticality.
Given equation (\ref{eq:mt}) it is not difficult to compute the mean
and the variance, which are given by\begin{eqnarray}
\overline{m} & = & \frac{1}{L}\sum_{k}\cos\left(\vartheta_{k}^{\left(2\right)}\right)\cos\left(\delta\vartheta_{k}\right)\label{eq:mave}\\
\overline{\Delta m^{2}} & = & \frac{1}{L^{2}}\sum_{k}\sin^{2}\left(\vartheta_{k}^{\left(2\right)}\right)\sin^{2}\left(\delta\vartheta_{k}\right).\label{eq:mvar}\end{eqnarray}
 Some comments are in order here. First fixing $h^{\left(1\right)},\, h^{\left(2\right)}$
the variance (figure \ref{fig:varm}) goes to zero as $L^{-1}$ and
not exponentially fast as was the case for the LE. Second, the spectral
weight associated to the frequency $\Lambda_{k}^{\left(2\right)}$
is $\sin\left(\vartheta_{k}^{\left(2\right)}\right)\sin\left(\delta\vartheta_{k}\right)$.
We observed that there are always many frequencies with large spectral
weight. In other words the spectral weight function is never peaked.

\begin{figure}
\noindent \begin{centering}
\includegraphics[width=7cm]{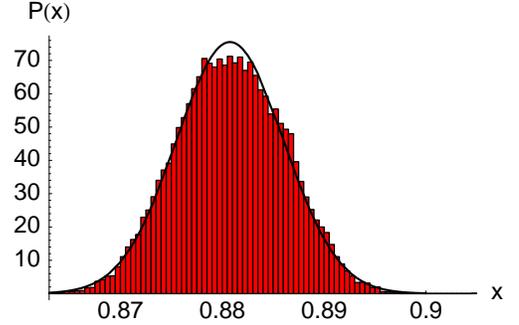} 
\par\end{centering}

\caption{(Color online) The probability distribution for the magnetization
has only Gaussian behavior. Here parameters are $L=40,\, h^{\left(1\right)}=0.9$,
$h^{\left(2\right)}=1.01$. The continuous line is a Gaussian with
mean and variance given by Eqns.~(\ref{eq:mave}) and (\ref{eq:mvar}).\label{fig:gauss-mag}}

\end{figure}

These comments suggest us to expect a Gaussian behavior for the probability
distribution function of the magnetization irrespective of the parameters
approaching critical values. This has indeed been observed (fig.~\ref{fig:gauss-mag}).

\section{Conclusions\label{sec:Conclusions}}

The unitary character of the dynamics of a closed quantum system implies
that whatever relevant notion of equilibration one might have has
to be a subtle one. In this paper we investigated the unitary equilibration
of a quantum system after a sudden change of its Hamiltonian parameter.
To this aim we used a prototypical time-dependent quantity: the Loschmidt
echo. We established how the global features of $\mathcal{L}$ depend
on the physical properties of the initial state preparation and on
those of the quench Hamiltonian. The central object of our analysis
is given by the long time probability distribution for $\mathcal{L}$:
$P(x)=\overline{\delta(\mathcal{L}(t)-x)}:=\lim_{T\to\infty}T^{-1}\int_{0}^{T}\delta\left(\mathcal{L}\left(t\right)-x\right)dt$.
Broadly speaking concentration phenomena for $P$ correspond to quantum
equilibration.

Here below for the reader's sake we summarize the main findings of
the paper 
\begin{itemize}
\item Resorting to a cumulant expansion we characterized the \char`\"{}short\char`\"{}
time behavior of $\mathcal{L}(t).$ Different regimes can be identified
depending on the most relevant scaling dimension of the quench Hamiltonian.
When the central limit theorem (CLT) applies one has a Gaussian decay
over a time scale $O(1)$ for gapped systems. For the critical case
the time-scale becomes $O\left(L^{\zeta}\right)$ in a small region
$\left|\delta h\right|\ll L^{-\left(d+\zeta-\Delta\right)}$. At critical
points the CLT can be violated and $\mathcal{L}(t)$ takes a universal
non-Gaussian form for sufficiently relevant perturbations. 
\item We discussed the general structure of the higher momenta of $P$ i.e.,
$\mu_{k}:=\overline{\mathcal{L}^{k}(t)}=\int x^{k}P(x)dx$ using the
so-called non-resonant hypothesis. We showed that all the $\mu_{k}$
are bounded by those corresponding to a Poissonian distribution i.e.,
$P(x)=\vartheta\left(x\right)\exp(-x/\overline{\mathcal{L}})/\overline{\mathcal{L}}$. 
\item Using exact results for the quantum Ising chain we rigorously analyzed
the interplay between the chain length $L$ and the averaging time
$T.$ In particular we showed how the limit $\lim_{T\to\infty}$ and
the thermodynamical one i.e., $\lim_{L\to\infty}$ do not commute.
While in finite systems the $\mathcal{L}(t)$ is an almost-periodic
function, in the thermodynamical limit $\mathcal{L}_{\infty}:=\lim_{t\to\infty}\mathcal{L}(t)$
exists and $P(x)\rightarrow\delta(x-\mathcal{L}_{\infty})$. We explicitly
computed $\mathcal{L}_{\infty}$ and the way it is asymptotically
approached for large $t.$ We gave a general closed form for the exact
$\mu_{k}$'s and compared with that obtained with the non-resonant
hypothesis 
\item For the quantum Ising chain we numerically investigated $P(x).$ We
identified three universal regimes a) An exponential one ($P$ is
Poissonian) when $L$ is the largest scale of the system b) A Gaussian
one for intermediate $L$ and initial state and quench parameters
close and off-critical c) A \char`\"{}Batman-hood\char`\"{} shape
for $P$ when the parameters are close to each other and close to
criticality. This result holds in the \emph{quasi-critical} region
$L|h^{(i)}-1|\ll1,\,(i=1,2).$ 
\item Finally, for the sake of the comparison of the Loschmidt echo with
a prototypical observable, we computed the time-dependent magnetization
after the quench and studied its long-time statistics. In this case
only a Gaussian regime appears to be reachable. 
\end{itemize}
We have shown that the Loschmidt echo encodes sophisticated information
about the quantum equilibration dynamics. For finite system vastly
different time-scales arise: short time relaxation is intertwined
with a complex a pattern of collapses and revivals and eventually
Poincare recurrences. Unveiling how these phenomena depend on spectral
properties of the underlying Hamiltonians, is one of the key challenges
in the way to understand emergent thermal behavior in closed quantum
systems.
\begin{acknowledgments}
The authors gratefully acknowledge discussions with H. Saleur and
A. Winter. Supported by NSF grants: PHY-803304,DMR-0804914 and European
project COQUIT FP7-ICT-2007-C, n. 233747. LCV wishes to thank S. Fortunato,
F. Radicchi and A. Lancichinetti for providing a result on the permutation
group and D. Burgarth for useful discussions. 
\end{acknowledgments}

\section*{Appendix}

\begin{figure}
\noindent \begin{centering}
\includegraphics[width=7cm]{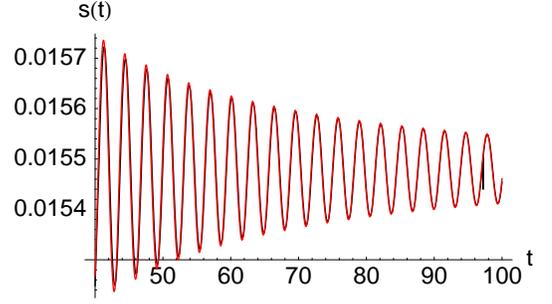} 
\par\end{centering}

\caption{Typical behavior of the function $s\left(t\right)$ (black line) in
the thermodynamic limit. The red line is the approximation given by
Eq.~(\ref{eq:LE-time_app}). Parameters are $h^{(1)}=1.3$, $h^{(2)}=2$.
\label{fig:losch-time}}

\end{figure}

\subsection*{Asymptotic of $s\left(t\right)$}

Here we want to compute the asymptotic of $s\left(t\right)$ for $t\to\infty$,
that is the integral:\[
s\left(t\right)=-\frac{1}{2\pi}\int_{0}^{\pi}\ln\left(1-\sin^{2}\left(\vartheta_{k}^{(1)}-\vartheta_{k}^{\left(2\right)}\right)\sin^{2}\left(\Lambda_{k}^{\left(2\right)}t/2\right)\right)dk\]
 We can go to energy integration setting $\Lambda_{k}^{\left(2\right)}=\omega$
\[
s\left(t\right)=-\frac{1}{2\pi}\int_{E_{m}}^{E_{M}}\ln\left[1-\alpha\left(\omega\right)\sin^{2}\left(\omega t/2\right)\right]\rho\left(\omega\right)d\omega.\]
 Where $E_{m}=2\min\left\{ \left|1+h^{(2)}\right|,\left|1-h^{(2)}\right|\right\} $
and $E_{M}=2\max\left\{ \left|1+h^{(2)}\right|,\left|1-h^{(2)}\right|\right\} $.
To be explicit:\begin{eqnarray*}
\rho\left(\omega\right) & = & \frac{2\omega}{\sqrt{\left(\omega^{2}-E_{m}^{2}\right)\left(E_{M}^{2}-\omega^{2}\right)}}\\
\alpha\left(\omega\right) & = & \frac{\left(\omega^{2}-E_{m}^{2}\right)\left(E_{M}^{2}-\omega^{2}\right)\left(h^{(2)}-h^{(1)}\right)^{2}}{4\left(h^{(2)}\right)^{2}\left[4\left(h^{(2)}-h^{(1)}\right)\left(1-h^{(1)}h^{(2)}\right)+h^{(1)}\omega^{2}\right]\omega^{2}}.\end{eqnarray*}
 Note that $\alpha\left(\omega\right)$ is zero at the band's edge,
positive otherwise (and smaller than 1 in modulus). Instead $\rho\left(\omega\right)$
has square root (van Hove) singularities at the band edges as a result
of the quadratic dispersion at those points (when $h^{(2)}\neq1$).
When $h^{(2)}=1$ the dispersion is linear at the bottom of the band
but still quadratic at the upper band edge, hence in this case only
the square root singularity at the upper band edge survives.

Then expand the logarithm into an infinite series. Using the Riemann-Lebesgue
lemma we can show that, \[
\lim_{t\to\infty}\int f\left(w\right)\left[\sin\left(wt\right)\right]^{2k}dw=2^{-2k}\left(\begin{array}{c}
2k\\
k\end{array}\right)\int f\left(w\right)d\omega,\]
 provided that $f$ is summable. The resulting series can be summed\[
-\sum_{k=1}^{\infty}\frac{x^{k}}{k}2^{-2k}\left(\begin{array}{c}
2k\\
k\end{array}\right)=2\ln\left(\frac{1+\sqrt{1-x}}{2}\right),\quad\mathrm{if}\,\,\left|x\right|<1.\]
 So finally\begin{eqnarray*}
\lim_{t\to\infty}s\left(t\right) & = & -\frac{1}{\pi}\int_{E_{m}}^{E_{M}}\ln\left[\frac{1+\sqrt{1-\alpha\left(\omega\right)}}{2}\right]\rho\left(\omega\right)d\omega\\
 & = & -\frac{1}{\pi}\int_{0}^{\pi}\ln\left[\frac{1+\sqrt{1-\alpha\left(k\right)}}{2}\right]dk.\end{eqnarray*}

To compute the first correction to the limit note that $\alpha\left(\omega\right)^{k}$
smoothen the singularity at the band edge, so that for the leading
correction we need only $k=1$ in the expansion of the logarithm.
The evaluation of the oscillating integral is done with a saddle point
technique. The result is\begin{eqnarray}
s\left(t\right) & \simeq & s\left(\infty\right)-\frac{1}{4\pi}\int_{E_{m}}^{E_{M}}\alpha\left(\omega\right)\rho\left(\omega\right)\cos\left(\omega t\right)d\omega\nonumber \\
 & \simeq & s\left(\infty\right)-\frac{A_{m}}{\left|t\right|^{3/2}}\cos\left(tE_{m}+\frac{3}{4}\pi\right)+\left(m\leftrightarrow M\right),\label{eq:LE-time_app}\end{eqnarray}
 with constants given by (we assumed here $h^{(2)}>0$ so that the
band minimum is $E_{m}=2\left|1-h^{(2)}\right|$)\begin{eqnarray*}
A_{m} & = & \frac{1}{16\sqrt{\pi}}\frac{\left(h^{(1)}-h^{(2)}\right)^{2}}{\left(1-h^{(1)}\right)^{2}\left(h^{\left(2\right)}\right)^{3/2}\sqrt{\left|1-h^{(2)}\right|}}\\
A_{M} & = & -\frac{1}{16\sqrt{\pi}}\frac{\left(h^{(1)}-h^{(2)}\right)^{2}}{\left(1+h^{(1)}\right)^{2}\left(h^{\left(2\right)}\right)^{3/2}\sqrt{\left|1+h^{(2)}\right|}}.\end{eqnarray*}
 The result (\ref{eq:LE-time_app}) should be the same as the square
modulus of Eq.~(12) in \cite{silva08}. However in \cite{silva08}
there appears only one frequency, corresponding to the lowest band
edge. The discrepancy probably arises from a continuum approximation
which discards the effect of the van Hove singularity present at the
upper band edge. As we have seen both terms give similar contributions.
In particular, even at criticality, the van Hove singularity at the
upper band edge survives.

In figure \ref{fig:losch-time} one can appreciate the validity of
the approximation (\ref{eq:LE-time_app}).

\bibliographystyle{apsrev}
\bibliography{ref_le}

\begin{thebibliography}{31}
\expandafter\ifx\csname natexlab\endcsname\relax\def\natexlab#1{#1}\fi
\expandafter\ifx\csname bibnamefont\endcsname\relax
  \def\bibnamefont#1{#1}\fi
\expandafter\ifx\csname bibfnamefont\endcsname\relax
  \def\bibfnamefont#1{#1}\fi
\expandafter\ifx\csname citenamefont\endcsname\relax
  \def\citenamefont#1{#1}\fi
\expandafter\ifx\csname url\endcsname\relax
  \def\url#1{\texttt{#1}}\fi
\expandafter\ifx\csname urlprefix\endcsname\relax\def\urlprefix{URL }\fi
\providecommand{\bibinfo}[2]{#2}
\providecommand{\eprint}[2][]{\url{#2}}

\bibitem[{\citenamefont{Calabrese and Cardy}(2006)}]{calabrese06}
\bibinfo{author}{\bibfnamefont{P.}~\bibnamefont{Calabrese}} \bibnamefont{and}
  \bibinfo{author}{\bibfnamefont{J.}~\bibnamefont{Cardy}},
  \bibinfo{journal}{\PRL} \textbf{\bibinfo{volume}{96}},
  \bibinfo{pages}{136801} (\bibinfo{year}{2006}).

\bibitem[{\citenamefont{Cazalilla}(2006)}]{cazalilla06}
\bibinfo{author}{\bibfnamefont{M.~A.} \bibnamefont{Cazalilla}},
  \bibinfo{journal}{\PRL} \textbf{\bibinfo{volume}{97}},
  \bibinfo{pages}{156403} (\bibinfo{year}{2006}).

\bibitem[{\citenamefont{Manmana et~al.}(2007)\citenamefont{Manmana, Wessel,
  Noack, and Muramatsu}}]{manmana07}
\bibinfo{author}{\bibfnamefont{S.~R.} \bibnamefont{Manmana}},
  \bibinfo{author}{\bibnamefont{Wessel}}, \bibinfo{author}{\bibfnamefont{R.~M.}
  \bibnamefont{Noack}}, \bibnamefont{and}
  \bibinfo{author}{\bibfnamefont{A.}~\bibnamefont{Muramatsu}},
  \bibinfo{journal}{\PRL} \textbf{\bibinfo{volume}{98}},
  \bibinfo{pages}{210405} (\bibinfo{year}{2007}).

\bibitem[{\citenamefont{Silva}(2008)}]{silva08}
\bibinfo{author}{\bibfnamefont{A.}~\bibnamefont{Silva}},
  \bibinfo{journal}{\PRL} \textbf{\bibinfo{volume}{101}},
  \bibinfo{pages}{120603} (\bibinfo{year}{2008}).

\bibitem[{\citenamefont{Kinoshita et~al.}(2006)\citenamefont{Kinoshita, Wegner,
  and Weiss}}]{newtons_cradle}
\bibinfo{author}{\bibfnamefont{T.}~\bibnamefont{Kinoshita}},
  \bibinfo{author}{\bibfnamefont{T.}~\bibnamefont{Wegner}}, \bibnamefont{and}
  \bibinfo{author}{\bibfnamefont{D.~S.} \bibnamefont{Weiss}},
  \bibinfo{journal}{Nature} \textbf{\bibinfo{volume}{440}},
  \bibinfo{pages}{900} (\bibinfo{year}{2006}).

\bibitem[{\citenamefont{Hoffenberth et~al.}(2007)\citenamefont{Hoffenberth,
  Lesanovsky, Fisher, Schumm, and Schmiedmayer}}]{hoffenberth07}
\bibinfo{author}{\bibfnamefont{S.}~\bibnamefont{Hoffenberth}},
  \bibinfo{author}{\bibfnamefont{I.}~\bibnamefont{Lesanovsky}},
  \bibinfo{author}{\bibfnamefont{B.}~\bibnamefont{Fisher}},
  \bibinfo{author}{\bibfnamefont{T.}~\bibnamefont{Schumm}}, \bibnamefont{and}
  \bibinfo{author}{\bibfnamefont{J.}~\bibnamefont{Schmiedmayer}},
  \bibinfo{journal}{Nature} \textbf{\bibinfo{volume}{449}},
  \bibinfo{pages}{324} (\bibinfo{year}{2007}).

\bibitem[{\citenamefont{Tasaki}(1998)}]{tasaki98}
\bibinfo{author}{\bibfnamefont{H.}~\bibnamefont{Tasaki}},
  \bibinfo{journal}{\PRL} \textbf{\bibinfo{volume}{80}}, \bibinfo{pages}{1373}
  (\bibinfo{year}{1998}).

\bibitem[{\citenamefont{Goldstein et~al.}(2006)\citenamefont{Goldstein,
  Leibowitz, Tumulka, and Zangh\'{i}}}]{goldsetin06}
\bibinfo{author}{\bibfnamefont{S.}~\bibnamefont{Goldstein}},
  \bibinfo{author}{\bibfnamefont{J.}~\bibnamefont{Leibowitz}},
  \bibinfo{author}{\bibfnamefont{R.}~\bibnamefont{Tumulka}}, \bibnamefont{and}
  \bibinfo{author}{\bibfnamefont{N.}~\bibnamefont{Zangh\'{i}}},
  \bibinfo{journal}{\PRL} \textbf{\bibinfo{volume}{96}},
  \bibinfo{pages}{050403} (\bibinfo{year}{2006}).

\bibitem[{\citenamefont{Reimann}(2007)}]{reimann2}
\bibinfo{author}{\bibfnamefont{P.}~\bibnamefont{Reimann}},
  \bibinfo{journal}{\PRL} \textbf{\bibinfo{volume}{99}},
  \bibinfo{pages}{160404} (\bibinfo{year}{2007}).

\bibitem[{\citenamefont{Reimann}(2008)}]{reimann08}
\bibinfo{author}{\bibfnamefont{P.}~\bibnamefont{Reimann}},
  \bibinfo{journal}{\PRL} \textbf{\bibinfo{volume}{101}},
  \bibinfo{pages}{190403} (\bibinfo{year}{2008}).

\bibitem[{\citenamefont{Popescu et~al.}(2006)\citenamefont{Popescu, Short, and
  Winter}}]{winter1}
\bibinfo{author}{\bibfnamefont{S.}~\bibnamefont{Popescu}},
  \bibinfo{author}{\bibfnamefont{A.~J.} \bibnamefont{Short}}, \bibnamefont{and}
  \bibinfo{author}{\bibfnamefont{A.}~\bibnamefont{Winter}},
  \bibinfo{journal}{Nature Physics} \textbf{\bibinfo{volume}{2}},
  \bibinfo{pages}{754} (\bibinfo{year}{2006}).

\bibitem[{\citenamefont{Linden et~al.}()\citenamefont{Linden, Popescu, Short,
  and Winter}}]{winter2}
\bibinfo{author}{\bibfnamefont{N.}~\bibnamefont{Linden}},
  \bibinfo{author}{\bibfnamefont{S.}~\bibnamefont{Popescu}},
  \bibinfo{author}{\bibfnamefont{A.~J.} \bibnamefont{Short}}, \bibnamefont{and}
  \bibinfo{author}{\bibfnamefont{A.}~\bibnamefont{Winter}},
  \bibinfo{note}{arXiv:0812.2385}.

\bibitem[{\citenamefont{Schotte and Schotte}(1969)}]{schotte69}
\bibinfo{author}{\bibfnamefont{K.~D.} \bibnamefont{Schotte}} \bibnamefont{and}
  \bibinfo{author}{\bibfnamefont{U.}~\bibnamefont{Schotte}},
  \bibinfo{journal}{\PR} \textbf{\bibinfo{volume}{182}}, \bibinfo{pages}{479}
  (\bibinfo{year}{1969}).

\bibitem[{\citenamefont{Prosen}(1998)}]{prosen98}
\bibinfo{author}{\bibfnamefont{T.}~\bibnamefont{Prosen}},
  \bibinfo{journal}{\PRL} \textbf{\bibinfo{volume}{80}}, \bibinfo{pages}{1808}
  (\bibinfo{year}{1998}).

\bibitem[{\citenamefont{Jalabert and Pastawski}(2001)}]{pastawski01}
\bibinfo{author}{\bibfnamefont{R.~A.} \bibnamefont{Jalabert}} \bibnamefont{and}
  \bibinfo{author}{\bibfnamefont{H.~M.} \bibnamefont{Pastawski}},
  \bibinfo{journal}{\PRL} \textbf{\bibinfo{volume}{86}}, \bibinfo{pages}{2490}
  (\bibinfo{year}{2001}).

\bibitem[{\citenamefont{Quan et~al.}(2006{\natexlab{a}})\citenamefont{Quan,
  Song, Liu, Zanardi, and Sun}}]{PZ-LE06}
\bibinfo{author}{\bibfnamefont{H.~T.} \bibnamefont{Quan}},
  \bibinfo{author}{\bibfnamefont{Z.}~\bibnamefont{Song}},
  \bibinfo{author}{\bibfnamefont{X.~F.} \bibnamefont{Liu}},
  \bibinfo{author}{\bibfnamefont{P.}~\bibnamefont{Zanardi}}, \bibnamefont{and}
  \bibinfo{author}{\bibfnamefont{C.~P.} \bibnamefont{Sun}},
  \bibinfo{journal}{\PRL} \textbf{\bibinfo{volume}{96}},
  \bibinfo{pages}{140604} (\bibinfo{year}{2006}{\natexlab{a}}).

\bibitem[{\citenamefont{Rossini
  et~al.}(2007{\natexlab{a}})\citenamefont{Rossini, Calarco, Giovannetti,
  Montangero, and Fazio}}]{rossini_base07}
\bibinfo{author}{\bibfnamefont{D.}~\bibnamefont{Rossini}},
  \bibinfo{author}{\bibfnamefont{T.}~\bibnamefont{Calarco}},
  \bibinfo{author}{\bibfnamefont{V.}~\bibnamefont{Giovannetti}},
  \bibinfo{author}{\bibfnamefont{S.}~\bibnamefont{Montangero}},
  \bibnamefont{and} \bibinfo{author}{\bibfnamefont{R.}~\bibnamefont{Fazio}},
  \bibinfo{journal}{J. Phys. A: Math. Theor.} \textbf{\bibinfo{volume}{40}},
  \bibinfo{pages}{8033} (\bibinfo{year}{2007}{\natexlab{a}}).

\bibitem[{\citenamefont{Rossini
  et~al.}(2007{\natexlab{b}})\citenamefont{Rossini, Calarco, Giovannetti,
  Montangero, and Fazio}}]{rossini07}
\bibinfo{author}{\bibfnamefont{D.}~\bibnamefont{Rossini}},
  \bibinfo{author}{\bibfnamefont{T.}~\bibnamefont{Calarco}},
  \bibinfo{author}{\bibfnamefont{V.}~\bibnamefont{Giovannetti}},
  \bibinfo{author}{\bibfnamefont{S.}~\bibnamefont{Montangero}},
  \bibnamefont{and} \bibinfo{author}{\bibfnamefont{R.}~\bibnamefont{Fazio}},
  \bibinfo{journal}{\PRA} \textbf{\bibinfo{volume}{75}},
  \bibinfo{pages}{032333} (\bibinfo{year}{2007}{\natexlab{b}}).

\bibitem[{\citenamefont{Zanardi et~al.}(2007)\citenamefont{Zanardi, Quan, Wang,
  and Sun}}]{PZ07}
\bibinfo{author}{\bibfnamefont{P.}~\bibnamefont{Zanardi}},
  \bibinfo{author}{\bibfnamefont{H.~T.} \bibnamefont{Quan}},
  \bibinfo{author}{\bibfnamefont{X.}~\bibnamefont{Wang}}, \bibnamefont{and}
  \bibinfo{author}{\bibfnamefont{C.~P.} \bibnamefont{Sun}},
  \bibinfo{journal}{\PRA} \textbf{\bibinfo{volume}{75}},
  \bibinfo{pages}{032109} (\bibinfo{year}{2007}).

\bibitem[{\citenamefont{Jarzynski}(1997)}]{jarzynski97}
\bibinfo{author}{\bibfnamefont{C.}~\bibnamefont{Jarzynski}},
  \bibinfo{journal}{\PRL} \textbf{\bibinfo{volume}{78}}, \bibinfo{pages}{2690}
  (\bibinfo{year}{1997}).

\bibitem[{\citenamefont{Kurchan}()}]{kurchan00}
\bibinfo{author}{\bibfnamefont{J.}~\bibnamefont{Kurchan}},
  \bibinfo{note}{arXive:cond-mat/0007360}.

\bibitem[{\citenamefont{Cramer et~al.}(2008)\citenamefont{Cramer, Flesch,
  McCulloch, Schollw\"{o}ck, and Eisert}}]{cramer08}
\bibinfo{author}{\bibfnamefont{M.}~\bibnamefont{Cramer}},
  \bibinfo{author}{\bibfnamefont{A.}~\bibnamefont{Flesch}},
  \bibinfo{author}{\bibfnamefont{I.~P.} \bibnamefont{McCulloch}},
  \bibinfo{author}{\bibfnamefont{U.}~\bibnamefont{Schollw\"{o}ck}},
  \bibnamefont{and} \bibinfo{author}{\bibfnamefont{J.}~\bibnamefont{Eisert}},
  \bibinfo{journal}{\PRL} \textbf{\bibinfo{volume}{101}},
  \bibinfo{pages}{063001} (\bibinfo{year}{2008}).

\bibitem[{not({\natexlab{a}})}]{nota}
\bibinfo{note}{Expanding up to second order in $\delta h$ around $h^{(2)}$,
  $c\left(\omega\right)$ can be well approximated by
  $c\left(\omega\right)=\frac{1}{8h_{2}^{2}\omega^{4}}\left(\omega^{2}-E_{m}^{%
2}\right)\left(E_{M}^{2}-\omega^{2}\right)\delta h^{2}+O\left(\delta
  h^{4}\right) $. This function has a flex at $\omega_\mathrm{flex} \simeq 3.6
  \left|1-h^{(2)}\right|$. Therefore the width of the peak is small compared to
  the bandwith roughly when $\left|1-h^{(2)}\right|\lesssim 10^{-1}$.}

\bibitem[{not({\natexlab{b}})}]{nota-lambda}
\bibinfo{note}{When $h^{(2)}$ is close to criticality, say $h^{(2)}= 1+\Delta
  h$, the band is approximately $\Lambda_k^{(2)}\simeq (1+\Delta
  h/2)|\sin(k/2)|$. The number of frequencies which fall in the peak is given
  by the $n$ satisfying $\Lambda^{(2)}_{k_n} = \omega_{\mathrm{flex}}$. At
  first order we obtain $n=O(L \Delta h)$, and so to have few freqencies in the
  peak we must have $L|h^{(2)}-1| \ll 1$.}

\bibitem[{\citenamefont{\mbox{Campos Venuti} and Zanardi}(2007)}]{LCV07}
\bibinfo{author}{\bibfnamefont{L.}~\bibnamefont{\mbox{Campos Venuti}}}
  \bibnamefont{and} \bibinfo{author}{\bibfnamefont{P.}~\bibnamefont{Zanardi}},
  \bibinfo{journal}{\PRL} \textbf{\bibinfo{volume}{99}},
  \bibinfo{pages}{095701} (\bibinfo{year}{2007}).

\bibitem[{\citenamefont{Privman et~al.}()\citenamefont{Privman, Hohenberg, and
  Aharony}}]{privman}
\bibinfo{author}{\bibfnamefont{V.}~\bibnamefont{Privman}},
  \bibinfo{author}{\bibfnamefont{P.}~\bibnamefont{Hohenberg}},
  \bibnamefont{and} \bibinfo{author}{\bibfnamefont{A.}~\bibnamefont{Aharony}},
  in \emph{\bibinfo{booktitle}{Phase Transition and Critical Phenomena}}
  (\bibinfo{publisher}{Academic Press}, \bibinfo{address}{London}, ????),
  vol.~\bibinfo{volume}{14}.

\bibitem[{\citenamefont{Aji and Goldenfeld}(2001)}]{aji01}
\bibinfo{author}{\bibfnamefont{V.}~\bibnamefont{Aji}} \bibnamefont{and}
  \bibinfo{author}{\bibfnamefont{N.}~\bibnamefont{Goldenfeld}},
  \bibinfo{journal}{\PRL} \textbf{\bibinfo{volume}{86}}, \bibinfo{pages}{1007}
  (\bibinfo{year}{2001}).

\bibitem[{\citenamefont{Lamacraft and Fendley}(2008)}]{fendley08}
\bibinfo{author}{\bibfnamefont{A.}~\bibnamefont{Lamacraft}} \bibnamefont{and}
  \bibinfo{author}{\bibfnamefont{P.}~\bibnamefont{Fendley}},
  \bibinfo{journal}{\PRL} \textbf{\bibinfo{volume}{100}},
  \bibinfo{pages}{165706} (\bibinfo{year}{2008}).

\bibitem[{not({\natexlab{c}})}]{nota-gap}
\bibinfo{note}{More precisely, assume the lattice is bipartite and the
  quasi-momenta satisfy some quantization in order to be roughly equally
  spaced: $k=2\pi n/L$. Then $\sum_{\mathbf{k}} (-1)^{\mathbf{k}}
  \Lambda_{\mathbf{k}}$ is the difference between two multi-dimensional Riemann
  sums. If $\Lambda_{\mathbf{k}}$ is analytic both these sums converge
  exponentially fast to their integral, and so their difference is
  exponentially small in $L$.}

\bibitem[{\citenamefont{Quan et~al.}(2006{\natexlab{b}})\citenamefont{Quan,
  Song, Liu, Zanardi, and Sun}}]{quan06}
\bibinfo{author}{\bibfnamefont{H.~T.} \bibnamefont{Quan}},
  \bibinfo{author}{\bibfnamefont{Z.}~\bibnamefont{Song}},
  \bibinfo{author}{\bibfnamefont{X.~F.} \bibnamefont{Liu}},
  \bibinfo{author}{\bibfnamefont{P.}~\bibnamefont{Zanardi}}, \bibnamefont{and}
  \bibinfo{author}{\bibfnamefont{C.~P.} \bibnamefont{Sun}},
  \bibinfo{journal}{\PRL} \textbf{\bibinfo{volume}{96}},
  \bibinfo{pages}{140604} (\bibinfo{year}{2006}{\natexlab{b}}).

\bibitem[{\citenamefont{Barouch et~al.}(1970)\citenamefont{Barouch, McCoy, and
  Dresden}}]{barouch70}
\bibinfo{author}{\bibfnamefont{E.}~\bibnamefont{Barouch}},
  \bibinfo{author}{\bibfnamefont{B.~M.} \bibnamefont{McCoy}}, \bibnamefont{and}
  \bibinfo{author}{\bibfnamefont{M.}~\bibnamefont{Dresden}},
  \bibinfo{journal}{\PRA} \textbf{\bibinfo{volume}{2}}, \bibinfo{pages}{1075}
  (\bibinfo{year}{1970}).

\end{thebibliography}

\end{document}